\documentclass[
showpacs,preprintnumbers,
nofootinbib,
 amsmath,amssymb,
 aps,
prd,
twocolumn
]{revtex4-1}

\usepackage[utf8]{inputenc}
\usepackage{epsf}
\usepackage{graphicx} 
\usepackage{amssymb}
\usepackage{dcolumn}
\usepackage{amsmath}
\usepackage{hyperref}
\usepackage{multirow}
\usepackage{booktabs}
\usepackage{siunitx}
\usepackage{color}
\newcommand{\kcmb}{\kappa_{\rm cmb}}

\def\thetaB{\mbox{\boldmath$\theta$}}

\def\ellB{\mbox{\boldmath$\ell$}}
\def\lsim{~\rlap{$<$}{\lower 1.0ex\hbox{$\sim$}}}
\def\gsim{~\rlap{$>$}{\lower 1.0ex\hbox{$\sim$}}}
\def\CB{\mbox{\boldmath${\rm C}$}}

\def\pB{\mbox{\boldmath$p$}}
\def\CB{\mbox{\boldmath${\rm C}$}}
\def\DB{\mbox{\boldmath$d$}}
\def\muB{\mbox{\boldmath$\mu$}}
\def\rhoB{\mbox{\boldmath$\rho$}}
\def\FB{\mbox{\boldmath$F$}}

\begin{document}

\title{CMB Lensing Beyond the Power Spectrum: Cosmological Constraints from the One-Point PDF and Peak Counts}

\author{Jia Liu$^{1,2}$}
\email{jia@astro.princeton.edu}

\author{J. Colin Hill$^{2}$}

\author{Blake D. Sherwin$^{3}$}

\author{Andrea Petri$^{4}$}

\author{Vanessa  B\"ohm$^{5}$}

\author{Zolt\'an Haiman$^{2,6}$}

\affiliation{ {$^1$ Department of Astrophysical Sciences, Princeton University, Princeton, NJ 08544, USA}} 
\affiliation{ {$^2$ Department of Astronomy, Columbia University, New York, NY 10027, USA}} 
\affiliation{ {$^3$ Berkeley Center for Cosmological Physics, LBL and Department of Physics, University of California, Berkeley, CA 94720, USA}}
\affiliation{ {$^4$ Department of Physics, Columbia University, New York, NY 10027, USA}} 
\affiliation{ {$^5$ Max-Planck-Institut f\"ur Astrophysik, D-85741 Garching, Germany}}
\affiliation{ {$^6$ Institute for Strings, Cosmology, and Astroparticle Physics (ISCAP), Columbia University, New York, NY 10027, USA}} 

\date{\today}

\begin{abstract}

Unprecedentedly precise cosmic microwave background (CMB) data are expected from ongoing and near-future CMB Stage-III and IV surveys, which will yield reconstructed CMB lensing maps with effective resolution approaching several arcminutes.  The small-scale CMB lensing fluctuations receive non-negligible contributions from nonlinear structure in the late-time density field.  These fluctuations are not fully characterized by traditional two-point statistics, such as the power spectrum.  Here, we use $N$-body ray-tracing simulations of CMB lensing maps to examine two higher-order statistics: the lensing convergence one-point probability distribution function (PDF) and peak counts.  We show that these statistics contain significant information not captured by the two-point function, and provide specific forecasts for the ongoing Stage-III Advanced Atacama Cosmology Telescope (AdvACT) experiment.  Considering only the temperature-based reconstruction estimator, we forecast 9$\sigma$ (PDF) and 6$\sigma$ (peaks) detections of these statistics with AdvACT.  Our simulation pipeline fully accounts for the non-Gaussianity of the lensing reconstruction noise, which is significant and cannot be neglected.  Combining the power spectrum, PDF, and peak counts for AdvACT will tighten cosmological constraints in the $\Omega_m$-$\sigma_8$ plane by $\approx 30\%$, compared to using the power spectrum alone.
\end{abstract}

\maketitle

\section{Introduction}\label{sec:intro}

After its first detection in cross-correlation nearly a decade ago~\cite{Smith2007,Hirata2008} and subsequent detection in auto-correlation five years ago~\cite{das2011,sherwin2011}, weak gravitational lensing of the cosmic microwave background (CMB) is now reaching maturity as a cosmological probe~\cite{Hanson2013,Das2013,PolarBear2014a,PolarBear2014b,BICEPKeck2016,Story2014,Ade2014,vanEngelen2014,vanEngelen2015,planck2015xv}. On their way to the Earth, CMB photons emitted at redshift $z=1100$ are deflected by the intervening matter, producing new correlations in maps of CMB temperature and polarization anisotropies.  Estimators based on these correlations can be applied to the observed anisotropy maps to reconstruct a noisy estimate of the CMB lensing potential~\cite{Zaldarriaga1998,Zaldarriaga1999,HuOkamoto2002,Okamoto2003}.  CMB lensing can probe fundamental physical quantities, such as the dark energy equation of state and neutrino masses, through its sensitivity to the geometry of the universe and the growth of structure (see Refs.~\cite{Lewis2006,Hanson2010} for a review). 

In this paper, we study the non-Gaussian information stored in CMB lensing observations. The Gaussian approximation to the density field breaks down due to nonlinear evolution on small scales at late times.  Thus, non-Gaussian statistics (i.e., statistics beyond the power spectrum) are necessary to capture the full information in the density field. Such work has been previously performed (theoretically and observationally) on weak gravitational lensing of galaxies, where galaxy shapes, instead of CMB temperature/polarization patterns, are distorted (hereafter ``galaxy lensing''). Several research groups have found independently that non-Gaussian statistics can tighten cosmological constraints when they are combined with the two-point correlation function or angular power spectrum.\footnote{For example, higher order moments~\cite{Bernardeau1997,Hui1999,vanWaerbeke2001, Takada2002,Zaldarriaga2003,Kilbinger2005,Petri2015}, 
three-point functions~\cite{Takada2003,Vafaei2010}, 
bispectra~\cite{Takada2004,DZ05,Sefusatti2006,Berge2010}, 
peak counts~\cite{Jain2000b,Marian2009,Maturi2010,Yang2011,Marian+2013,Liu2015,Liu2014b,Lin&Kilbinger2015a,Lin&Kilbinger2015b,Kacprzak2016}, 
Minkowski functionals~\cite{Kratochvil2012,Shirasakiyoshida2014,Petri2013,Petri2015}, and Gaussianized power spectrum~\cite{Neyrinck2009,Neyrinck2014,Yu2012}.}  Such non-Gaussian statistics have also been applied in the CMB context to the Sunyaev-Zel'dovich signal, including higher-order moments~\cite{Wilson2012,Hill2013,Planck2013tSZ,Planck2015tSZ}, the bispectrum~\cite{Bhattacharya2012,Crawford2014,Planck2013tSZ,Planck2015tSZ}, and the one-point probability distribution function (PDF)~\cite{Hill2014b,Planck2013tSZ,Planck2015tSZ}.  In all cases, substantial non-Gaussian information was found, yielding improved cosmological constraints.

The motivation to study non-Gaussian statistics of CMB lensing maps is three-fold.  First, the CMB lensing kernel is sensitive to structures at high redshift ($z\approx2.0$, compared to $z\approx0.4$ for typical galaxy lensing samples); hence CMB lensing non-Gaussian statistics probe early nonlinearity that is beyond the reach of galaxy surveys. Second, CMB lensing does not suffer from some challenging systematics that are relevant to galaxy lensing, including intrinsic alignments of galaxies, photometric redshift uncertainties, and shape measurement biases. Therefore, a combined analysis of galaxy lensing and CMB lensing will be useful to build a tomographic outlook on nonlinear structure evolution, as well as to calibrate systematics in both galaxy and CMB lensing surveys~\cite{Liu2016,Baxter2016,Schaan2016,Singh2016,Nicola2016}.  Finally, CMB lensing measurements have recently entered a regime of sufficient sensitivity and resolution to detect the (stacked) lensing signals of halos~\cite{Madhavacheril2014,Baxter2016,Planck2015cluster}.  This suggests that statistics sensitive to the nonlinear growth of structure, i.e., non-Gaussian statistics, will also soon be detectable.  We demonstrate below that this is indeed the case, taking as a reference experiment the ongoing Advanced Atacama Cosmology Telescope (AdvACT) survey~\cite{Henderson2016}.

Non-Gaussian aspects of the CMB lensing field have recently attracted attention, both as a potential signal and a source of bias in CMB lensing power spectrum estimates.  Considering the lensing non-Gaussianity as a signal, a recent analytical study of the CMB lensing bispectrum by Ref.~\cite{Namikawa2016} forecasted its detectability to be 40$\sigma$ with a CMB Stage-IV experiment.  Ref.~\cite{Bohm2016} performed the first calculation of the bias induced in CMB lensing power spectrum estimates by the lensing bispectrum, finding non-negligible biases for Stage-III and IV CMB experiments.  Refs.~\cite{Pratten2016} and~\cite{Marozzi2016} considered CMB lensing effects arising from the breakdown of the Born approximation, with the former study finding that post-Born terms substantially alter the predicted CMB lensing bispectrum, compared to the contributions from nonlinear structure formation alone.  We emphasize that the $N$-body ray-tracing simulations used in this work naturally capture such effects --- we do not use the Born approximation.  However, we consider only the lensing potential $\phi$ or convergence $\kappa$ here (related by $\kappa = -\nabla^2 \phi/2$), leaving a treatment of the curl potential or image rotation for future work (Ref.~\cite{Pratten2016} has demonstrated that the curl potential possesses non-trivial higher-order statistics).  In a follow-up paper, the simulations described here are used to more precisely characterize CMB lensing power spectrum biases arising from the bispectrum and higher-order correlations \cite{Sherwin2016}.

We consider the non-Gaussianity in the CMB lensing field as a potential signal.  We use a suite of 46 $N$-body ray-tracing simulations to investigate two non-Gaussian statistics applied to CMB lensing convergence maps --- the one-point PDF and peak counts. We examine the deviation of the convergence PDF and peak counts from those of Gaussian random fields. We then quantify the power of these statistics to constrain cosmological models, compared with using the power spectrum alone.

The paper is structured as follows. We first introduce CMB lensing in Sec.~\ref{sec:formalism}. We then describe our simulation pipeline in Sec.~\ref{sec:sim} and analysis procedures in Sec.~\ref{sec:analysis}. We show our results for the power spectrum, PDF, peak counts, and the derived cosmological constraints in Sec.~\ref{sec:results}. We conclude in Sec.~\ref{sec:conclude}.

\section{CMB lensing formalism}\label{sec:formalism}

To lowest order, the lensing convergence ($\kappa$) is a weighted projection of the three-dimensional matter overdensity $\delta=\delta\rho/\bar{\rho}$ along the line of sight,

\begin{equation}
\label{eq.kappadef}
\kappa(\thetaB) = \int_0^{\infty} dz W(z) \delta(\chi(z)\thetaB, z),
\end{equation}
where $\chi(z)$ is the comoving distance and the kernel $W(z)$ indicates the lensing strength at redshift $z$ for sources with a redshift distribution $p(z_s)=dn(z_s)/dz$. For CMB lensing, there is only one source plane at the last scattering surface $z_\star=1100$; therefore, $p(z_s)=\delta_D(z_s-z_\star)$, where $\delta_D$ is the Dirac delta function. For a flat universe, the CMB lensing kernel is

\begin{eqnarray}
W^{\kcmb}(z) &=&  \frac{3}{2}\Omega_{m}H_0^2  \frac{(1+z)}{H(z)} \frac{\chi(z)}{c} \nonumber\\ 
&\times&  \frac{\chi(z_\star)-\chi(z)}{\chi(z_\star)}.
\end{eqnarray}
where $\Omega_{m}$ is the matter density as a fraction of the critical density at $z=0$, $H(z)$ is the Hubble parameter at redshift $z$, with a present-day value $H_0$, and $c$ is the speed of light. $W^{\kcmb}(z)$ peaks at $z\approx2$ for canonical cosmological parameters ($\Omega_{m}\approx0.3$ and $H_0\approx70$ km/s/Mpc,~\cite{planck2015xiii}).  Note that Eq.~(\ref{eq.kappadef}) assumes the Born approximation, but our simulation approach described below does not --- we implement full ray-tracing to calculate $\kappa$.

\section{Simulations}\label{sec:sim}

Our simulation procedure includes five main steps: (1) the design (parameter sampling) of cosmological models, (2) $N$-body simulations with Gadget-2,\footnote{\url{http://wwwmpa.mpa-garching.mpg.de/gadget/}} (3) ray-tracing from $z=0$ to $z=1100$ to obtain (noiseless) convergence maps using the Python code LensTools~\cite{Petri2016},\footnote{\url{https://pypi.python.org/pypi/lenstools/}} (4) lensing simulated CMB temperature maps by the ray-traced convergence field, and (5) reconstructing (noisy) convergence maps from the CMB temperature maps after including noise and beam effects.

\subsection{Simulation design}

We use an irregular grid to sample parameters in the $\Omega_m$-$\sigma_8$ plane, within the range of $\Omega_m \in [0.15, 0.7]$ and $\sigma_8 \in [0.5, 1.0]$, where $\sigma_8$ is the rms amplitude of linear density fluctuations on a scale of 8 Mpc/$h$ at $z=0$. An optimized irregular grid has a smaller average distance between neighboring points than a regular grid, and no parameters are duplicated.  Hence, it samples the parameter space more efficiently. The procedure to optimize our sampling is described in detail in Ref.~\cite{Petri2015}. 

\begin{figure}
\begin{center}
\includegraphics[width=0.48\textwidth]{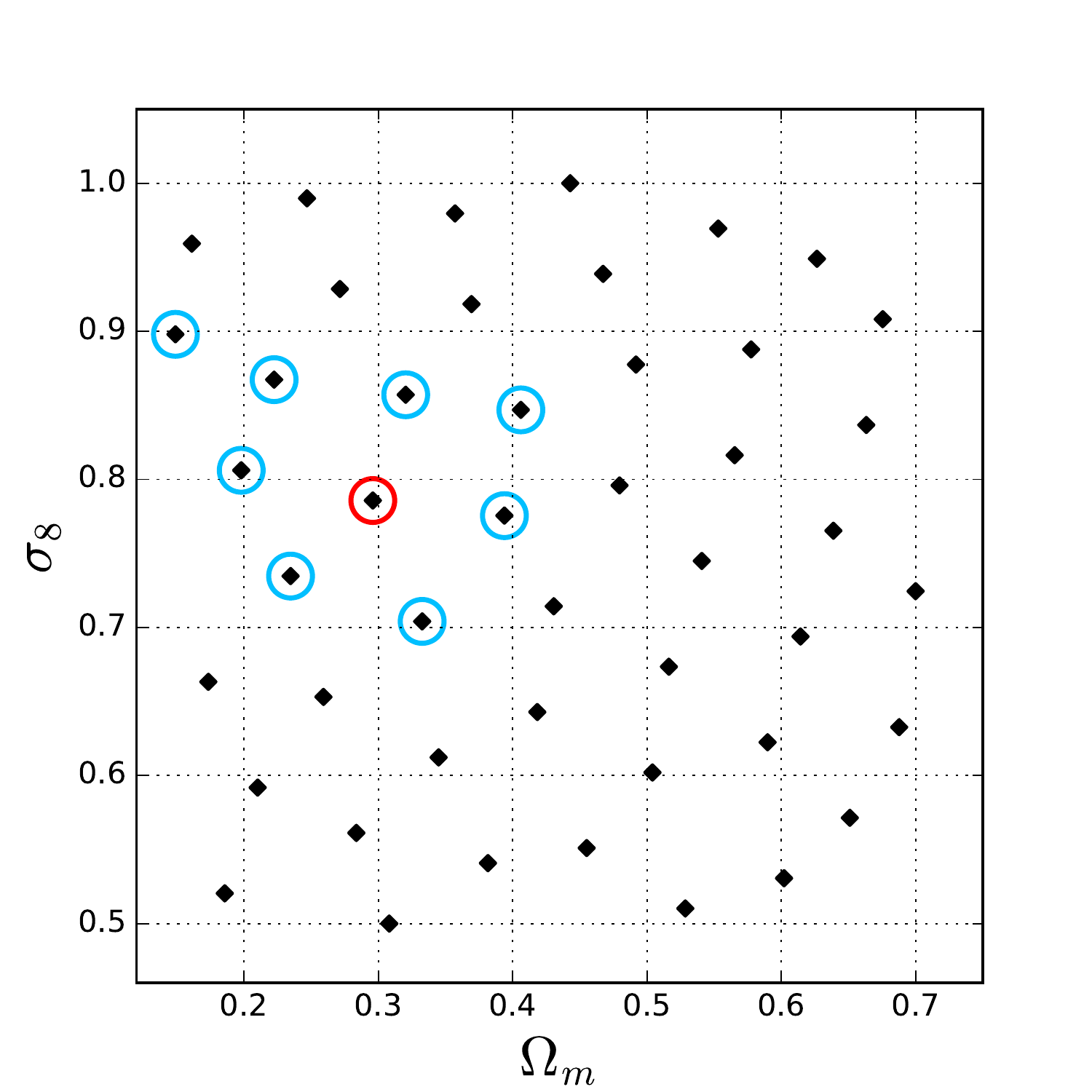}
\end{center}
\caption{\label{fig:design} The design of cosmological parameters used in our simulations (46 models in total). The fiducial cosmology ($\Omega_m=0.296, \sigma_8=0.786$) is circled in red. The models for which AdvACT-like lensing reconstruction is performed are circled in blue.  Other cosmological parameters are fixed at $H_0=72$ km/s/Mpc,  $w=-1$, $n_s=0.96$, and $\Omega_b=0.046$.}
\end{figure}

The 46 cosmological models sampled are shown in Fig.~\ref{fig:design}. Other cosmological parameters are held fixed, with $H_0=72$ km/s/Mpc, dark energy equation of state $w=-1$, spectral index $n_s=0.96$, and baryon density $\Omega_b=0.046$.  The design can be improved in the future by posterior sampling, where we first run only a few models to generate a low-resolution probability plane, and then sample more densely in the high-probability region.

We select the model that is closest to the standard concordance values of the cosmological parameters (e.g.,~\cite{planck2015xiii}) as our fiducial model, with $\Omega_m=0.296$ and $\sigma_8=0.786$. We create two sets of realizations for this model, one for covariance matrix estimation, and another one for parameter interpolation. This fiducial model is circled in red in Fig.~\ref{fig:design}.

\subsection{$N$-body simulation and ray-tracing}\label{sec:nbody}

We use the public code Gadget-2 to run $N$-body simulations with $N_{\rm particles}=1024^3$ and box size = 600 Mpc/$h$ (corresponding to a mass resolution of $1.4\times10^{10} M_\odot/h$). To initialize each simulation, we first obtain the linear matter power spectrum with the Einstein-Boltzmann code CAMB.\footnote{\url{http://camb.info/}} The power spectrum is then fed into the initial condition generator N-GenIC, which generates initial snapshots (the input of Gadget-2) of particle positions at $z=100$. The $N$-body simulation is then run from $z=100$ to $z=0$, and we record snapshots at every 144 Mpc$/h$ in comoving distance between $z\approx45$ and $z=0$. The choice of $z\approx45$ is determined by requiring that the redshift range covers 99\% of the $W^{\kappa_{cmb}}D(z)$ kernel, where we use the linear growth factor $D(z)\sim 1/(1+z)$. 

We then use the Python code LensTools~\cite{Petri2016} to generate CMB lensing convergence maps. We first slice the simulation boxes to create potential planes (3 planes per box,  200 Mpc/$h$ in thickness), where particle density is converted into gravitational potential using the Poisson equation. We track the trajectories of 4096$^2$ light rays from $z=0$ to $z=1100$, where the deflection angle and convergence are calculated at each potential plane.  This procedure automatically captures so-called ``post-Born'' effects, as we never assume that the deflection angle is small or that the light rays follow unperturbed geodesics.\footnote{While the number of potential planes could be a limiting factor in our sensitivity to these effects, we note that our procedure uses $\approx 40$-70 planes for each ray-tracing calculation (depending on the cosmology), which closely matches the typical number of lensing deflections experienced by a CMB photon.}  Finally, we create 1,000 convergence map realizations for each cosmology by randomly rotating/shifting the potential planes~\cite{Petri2016b}.  For the fiducial cosmology only, we generate 10,000 realizations for the purpose of estimating the covariance matrix.  The convergence maps are 2048$^2$ pixels and 12.25 deg$^2$ in size, with square pixels of side length 0.1025 arcmin. The maps generated at this step correspond to the physical lensing convergence field only, i.e., they have no noise from CMB lensing reconstruction. Therefore, they are labeled as ``noiseless'' in the following sections and figures.

\begin{figure}
\begin{center}
\includegraphics[width=0.48\textwidth]{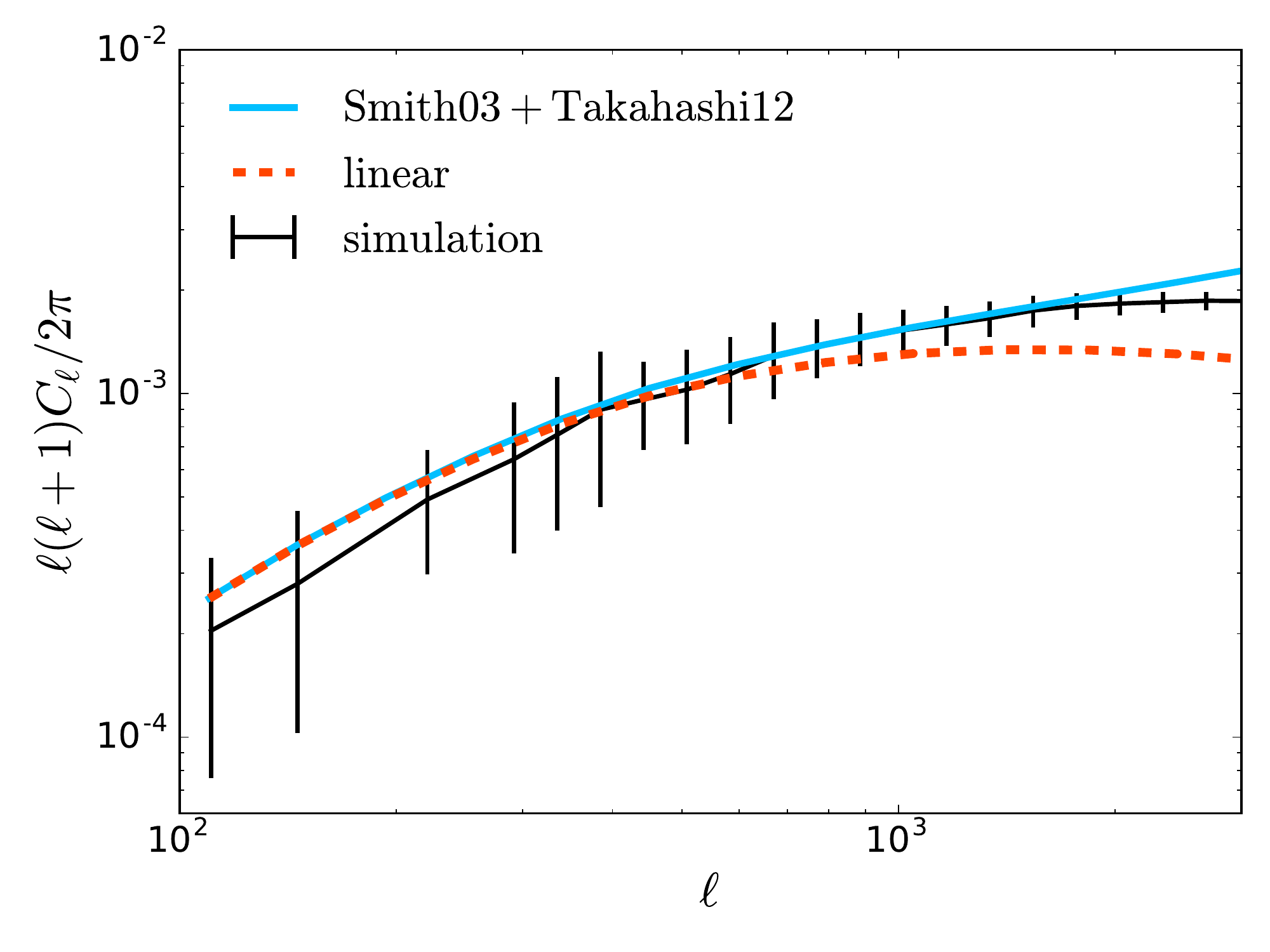}
\end{center}
\caption{\label{fig:theory_ps} Comparison of the CMB lensing convergence power spectrum from the HaloFit model and that from our simulation (1024$^3$ particles, box size 600 Mpc/$h$, map size 12.25 deg$^2$), for our fiducial cosmology. We also show the prediction from linear theory. Error bars are the standard deviation of 10,000 realizations.}
\end{figure}

We test the power spectra from our simulated maps against standard theoretical predictions. Fig.~\ref{fig:theory_ps} shows the power spectrum from our simulated maps versus that from the HaloFit model~\cite{Smith2003, Takahashi2012} for our fiducial cosmology.  We also show the linear-theory prediction, which deviates from the nonlinear HaloFit result at $\ell \gtrsim 700$.  The simulation error bars are estimated using the standard deviation of 10,000 realizations. The simulated and (nonlinear) theoretical results are consistent within the error bars for multipoles $\ell<2,000$, which is sufficient for this work, as current and near-future CMB lensing surveys are limited to roughly this $\ell$ range due to their beam size and noise level (the filtering applied in our analysis below effectively removes all information on smaller angular scales).  We find similar consistency between theory and simulation for the other 45 simulated models.  We test the impact of particle resolution using a smaller box of 300 Mpc/$h$, while keeping the same number of particles (i.e. 8 times higher resolution), and obtain excellent agreement at scales up to $\ell=3,000$.
The lack of power on large angular scales is due to the limited size of our convergence maps, while the missing power on small scales is due to our particle resolution.  On very small scales ($\ell \gtrsim 5 \times 10^4$), excess power due to finite-pixelization shot noise arises, but this effect is negligible on the scales considered in our analysis.

\subsection{CMB lensing reconstruction}\label{sec:recon}

\begin{figure*}
\begin{center}
\includegraphics[width=\textwidth]{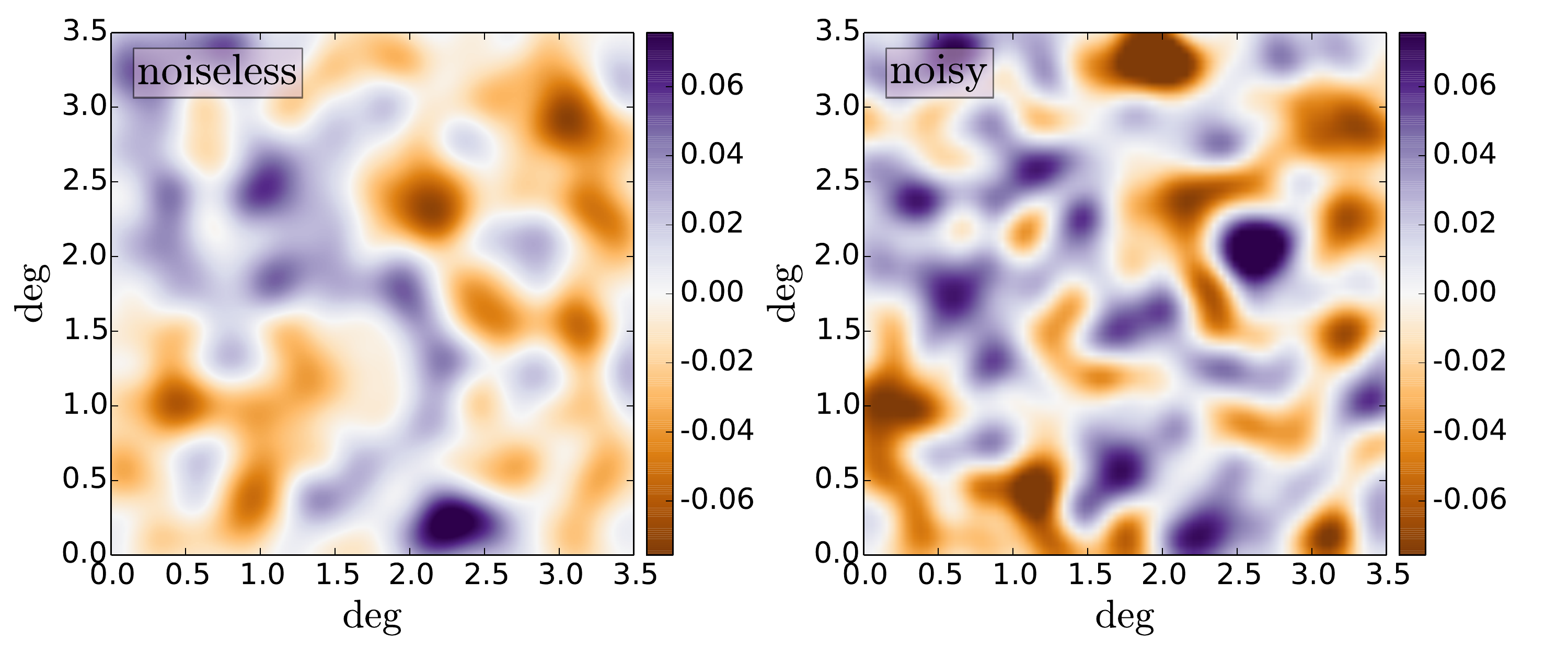}
\end{center}
\caption{\label{fig:sample_maps} Example of a simulated convergence map in the fiducial cosmology, before (left panel) and after (right panel) reconstruction, assuming AdvACT-like experimental specifications (6 $\mu$K-arcmin noise, $1.4$ arcmin beam).  The maps are smoothed with a FWHM$=8$ arcmin Gaussian window function for visual purposes. The color scale is set to $\pm3\sigma_{\kappa}^{\rm noiseless} $, where $\sigma_\kappa^{\rm noiseless} = 0.025$ is the rms of the noiseless map (in comparison, the rms of the noisy map is $\sigma_\kappa^{\rm noisy} = 0.034$).}
\end{figure*}

In order to obtain CMB lensing convergence maps with realistic noise properties, we generate lensed CMB temperature maps and reconstruct noisy estimates of the convergence field.  First, we generate Gaussian random field CMB temperature maps based on a $\Lambda$CDM concordance model temperature power spectrum computed with CAMB.  We compute deflection field maps from the ray-traced convergence maps described in the previous sub-section, after applying a filter that removes power in the convergence maps above $
\ell \approx 4,000$.\footnote{We find that this filter is necessary for numerical stability (and also because our simulated $\kappa$ maps do not recover all structure on these small scales, as seen in Fig.~\ref{fig:theory_ps}), but our results are unchanged for moderate perturbations to the filter scale.}  These deflection maps are then used to lens the simulated primary CMB temperature maps.  The lensing simulation procedure is described in detail in Ref.~\cite{Louis2013}.

After obtaining the lensed temperature maps, we apply instrumental effects consistent with specifications for the ongoing AdvACT survey~\cite{Henderson2016}.  In particular, the maps are smoothed with a FWHM $=1.4$ arcmin beam, and Gaussian white noise of amplitude 6$\mu$K-arcmin is then added.

We subsequently perform lensing reconstruction on these beam-convolved, noisy temperature maps using the quadratic estimator of Ref.~\cite{HuOkamoto2002}, but with the replacement of unlensed with lensed CMB temperature power spectra in the filters, which gives an unbiased reconstruction to higher order \cite{Hanson2010}.  The final result is a noisy estimate of the CMB lensing convergence field, with 1,000 realizations for each cosmological model (10,000 for the fiducial model).

We consider only temperature-based reconstruction in this work, leaving polarization estimators for future consideration.  The temperature estimator is still expected to contribute more significantly than the polarization to the signal-to-noise for Stage-III CMB experiments like AdvACT, but polarization will dominate for Stage-IV (via $EB$ reconstruction).  For the AdvACT-like experiment considered here, including polarization would increase the predicted signal-to-noise on the lensing power spectrum by $\approx 35$\%.  More importantly, polarization reconstruction allows the lensing field to be mapped out to smaller scales than temperature reconstruction~\cite{HuOkamoto2002}, and is more immune to foreground-related biases at high-$\ell$~\cite{vanEngelen2014b}.  Thus, it could prove extremely useful for higher-order CMB lensing statistics, which are sourced by non-Gaussian structure on small scales.  Clearly these points are worthy of future analysis, but we restrict this work to temperature reconstruction for simplicity.

In addition to the fiducial model, we select the nearest eight points in the sampled parameter space (points circled in blue in Fig.~\ref{fig:design}) for the reconstruction analysis. We determine this selection by first reconstructing the nearest models in parameter space, and then broadening the sampled points until the interpolation is stable and the forecasted contours (see Sec.~\ref{sec:constraints}) are converged for AdvACT-level noise.  At this noise level, the other points in model space are sufficiently distant to contribute negligibly to the forecasted contours.  In total, nine models are used to derive parameter constraints from the reconstructed, noisy maps.  For completeness, we perform a similar convergence test using forecasted constraints from the noiseless maps, finding excellent agreement between contours derived using all 46 models and using only these nine models.

In Fig.~\ref{fig:sample_maps}, we show an example of a convergence map from the fiducial cosmology before (``noiseless'') and after (``noisy'') reconstruction. Prominent structures seen in the noiseless maps remain obvious in the reconstructed, noisy maps.

\subsection{Gaussian random field}

We also reconstruct a set of Gaussian random fields (GRF) in the fiducial model.  We generate a set of GRFs using the average power spectrum of the noiseless $\kappa$ maps.  We then lens simulated CMB maps using these GRFs, following the same procedure as outlined above, and subsequently perform lensing reconstruction, just as for the reconstructed $N$-body $\kappa$ maps. These noisy GRF-only reconstructions allow us to examine the effect of reconstruction (in particular the non-Gaussianity of the reconstruction noise itself), as well as to determine the level of non-Gaussianity in the noisy $\kappa$ maps.

\subsection{Interpolation}

\begin{figure}
\begin{center}
\includegraphics[width=0.48\textwidth]{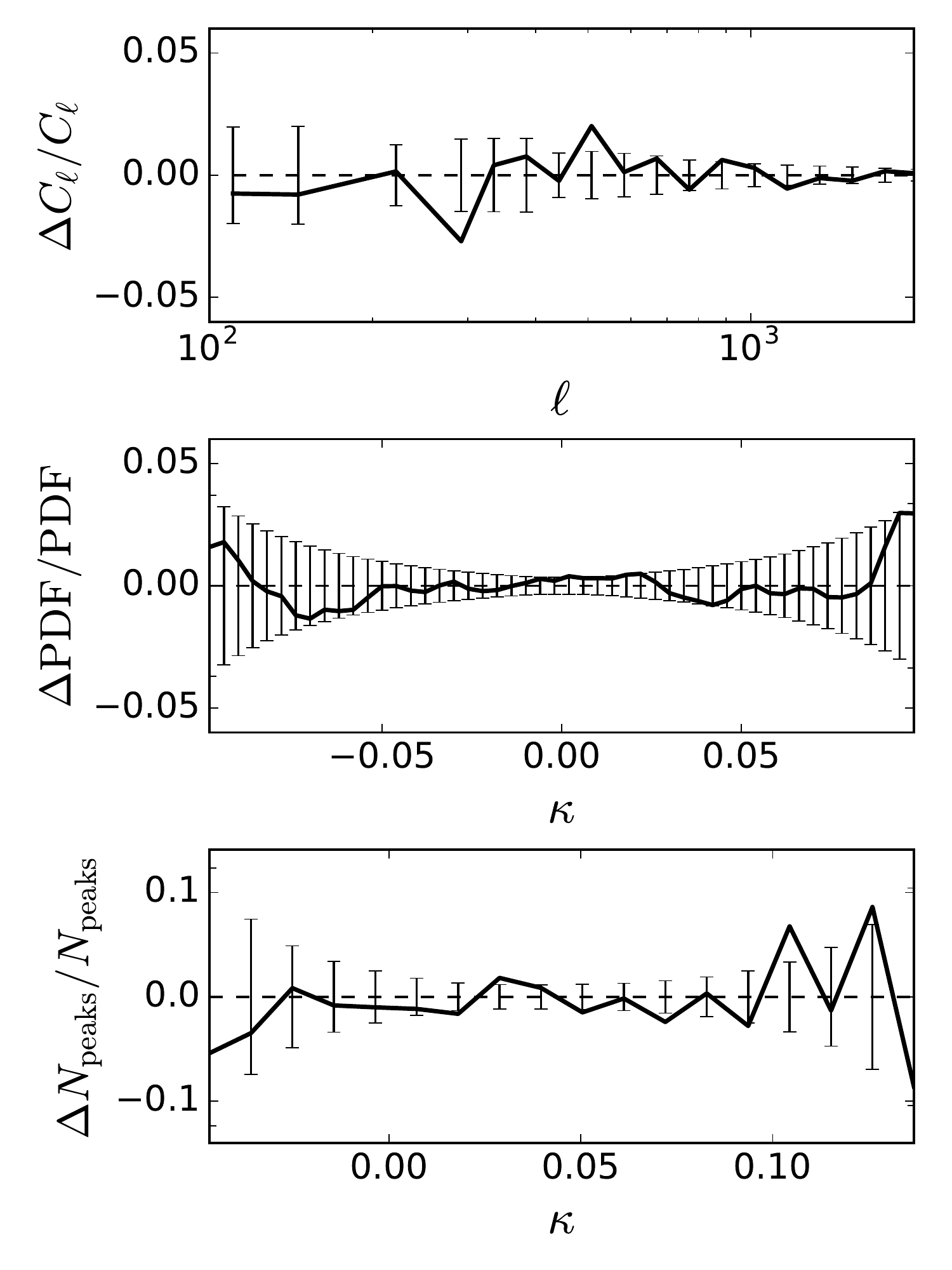}
\end{center}
\caption{\label{fig:interp} Fractional differences between interpolated and ``true'' results for the fiducial power spectrum (top), PDF (middle), and peak counts (bottom).  Here, we have built the interpolator using results for the other 45 cosmologies, and then compared the interpolated prediction at the fiducial parameter values to the actual simulated results for the fiducial cosmology. The error bars are scaled by $1/\sqrt{N_{\rm sims}}$, where the number of simulations $N_{\rm sims}=1,000$.  The agreement for all three statistics is excellent.}
\end{figure}

To build a model at points where we do not have simulations, we interpolate from the simulated points in parameter space using the Clough-Tocher interpolation scheme~\cite{alfeld1984,farin1986}, which triangulates the input points and then minimizes the curvature of the interpolating surface; the interpolated points are guaranteed to be continuously differentiable. In Fig.~\ref{fig:interp}, we show a test of the interpolation using the noiseless $\kappa$ maps: we build the interpolator using all of the simulated cosmologies except for the fiducial model (i.e., 45 cosmologies), and then compare the interpolated results at the fiducial parameter values with the true, simulated results for that cosmology.  The agreement for all three statistics is excellent, with deviations $\lesssim$ few percent (and well within the statistical precision).  Finally, to check the robustness of the interpolation scheme, we also run our analysis using linear interpolation, and obtain consistent results.\footnote{Due to our limited number of models, linear interpolation is slightly more vulnerable to sampling artifacts than the Clough-Tocher method, because the linear method only utilizes the nearest points in parameter space.  The Clough-Tocher method also uses the derivative information. Therefore, we choose Clough-Tocher for our analysis.}

\section{Analysis}\label{sec:analysis}

In this section, we describe the analysis of the simulated CMB lensing maps, including the computation of the power spectrum, peak counts, and PDF, and the likelihood estimation for cosmological parameters.  These procedures are applied in the same way to the noiseless and noisy (reconstructed) maps.

\subsection{Power spectrum,  PDF, and peak counts}

To compute the power spectrum, we first estimate the two-dimensional (2D) power spectrum of CMB lensing maps ($M_{\kappa}$) using
\begin{eqnarray}
\label{eq: ps2d}
C^{\kappa \kappa}(\ellB) = \hat M_{\kappa}(\ellB)^*\hat M_{\kappa}(\ellB) \,,
\end{eqnarray}
where $\ellB$ is the 2D multipole with components $\ell_1$ and $\ell_2$, $\hat M_{\kappa}$ is the Fourier transform of $M_{\kappa}$, and the asterisk denotes complex conjugation.  We then average over all the pixels within each $|\ellB|\in[\ell-\Delta\ell, \ell+\Delta\ell)$ bin, for 20 log-spaced bins in the range of $100<\ell<2,000$, to obtain the one-dimensional power spectrum.

The one-point PDF is the number of pixels with values between [$\kappa-\Delta\kappa$, $\kappa+\Delta\kappa$) as a function of $\kappa$. We use 50 linear bins with edges listed in Table~\ref{tab: bins}, and normalize the resulting PDF such that its integral is unity. The PDF is a simple observable (a histogram of the data), but captures the amplitude of all (zero-lag) higher-order moments in the map.  Thus, it provides a potentially powerful characterization of the non-Gaussian information.

Peaks are defined as local maxima in a $\kappa$ map. In a pixelized map, they are pixels with values higher than the surrounding 8 (square) pixels. Similar to cluster counts, peak counts are sensitive to the most nonlinear structures in the Universe. For galaxy lensing, they have been found to associate with halos along the line of sight both with simulations~\cite{Yang2011} and observations~\cite{LiuHaiman2016}. We record peaks on smoothed $\kappa$ maps, in 25 linearly spaced bins with edges listed in Table~\ref{tab: bins}.

\begin{table}
\begin{tabular}{|c|c|c|}
\hline
Smoothing scale &       PDF bins edges  &       Peak counts bin edges           \\
(arcmin)                        &        (50 linear bins)               &       (25 linear bins)                \\
\hline                                                                                                                  
0.5 (noiseless)         &    [-0.50, +0.50]     &       [-0.18, +0.36]  \\
1.0 (noiseless)         &    [-0.22, +0.22]     &       [-0.15, +0.30]  \\
2.0 (noiseless)         &    [-0.18, +0.18]     &       [-0.12, +0.24]  \\
5.0 (noiseless)         &    [-0.10, +0.10]     &       [-0.09, +0.18]  \\
8.0 (noiseless)         &    [-0.08, +0.08]     &       [-0.06, +0.12]  \\
1.0, 5.0, 8.0 (noisy)               &    [-0.12, +0.12]     &       [-0.06, +0.14]  \\
\hline
\end{tabular}
\caption[]{\label{tab: bins} PDF and peak counts bin edges for each smoothing scale (the full-width-half-maximum of the Gaussian smoothing kernel applied to the maps).}
\end{table}

\subsection{Cosmological constraints}

We estimate cosmological parameter confidence level (C.L.) contours assuming a constant (cosmology-independent) covariance and Gaussian likelihood,
\begin{align}
P (\DB | \pB) = \frac{1}{2\pi|\CB|^{1/2}} \exp\left[-\frac{1}{2}(\DB-\muB)\CB^{-1}(\DB-\muB)\right],
\end{align}
where $\DB$ is the data array, $\pB$ is the input parameter array, $\muB=\muB(\pB)$ is the interpolated model, and $\CB$ is the covariance matrix estimated using the fiducial cosmology, with determinant $|\CB|$. The correction factor for an unbiased inverse covariance estimator~\cite{dietrich2010} is negligible in our case, with $(N_{\rm sims}-N_{\rm bins}-2)/(N_{\rm sims}-1) = 0.99$ for $N_{\rm sims} =10,000$ and $N_{\rm bins}=95$.  We leave an investigation of the impact of cosmology-dependent covariance matrices and a non-Gaussian likelihood for future work.

Due to the limited size of our simulated maps, we must rescale the final error contour by a ratio ($r_{\rm sky}$) of simulated map size  (12.25 deg$^2$) to the survey coverage (20,000 deg$^2$ for AdvACT). Two methods allow us to achieve this --- rescaling the covariance matrix by $r_{\rm sky}$ before computing the likelihood plane, or rescaling the final C.L. contour by $r_{\rm sky}$. These two methods yield consistent results. In our final analysis, we choose the former method.

\section{Results}\label{sec:results}

\subsection{Non-Gaussianity in noiseless maps}\label{sec:non-gauss}

\begin{figure*}
\begin{center}
\includegraphics[width=0.48\textwidth]{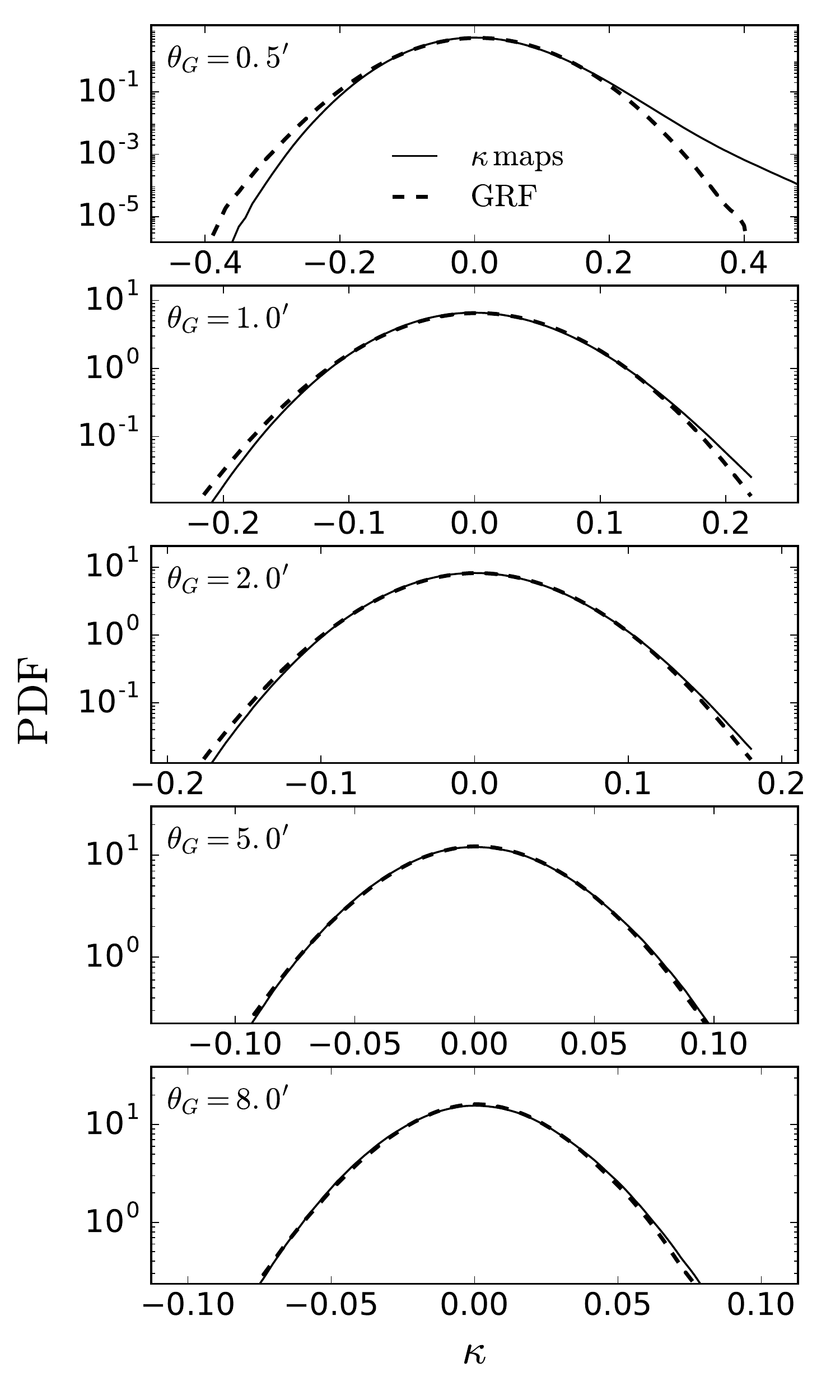}
\includegraphics[width=0.48\textwidth]{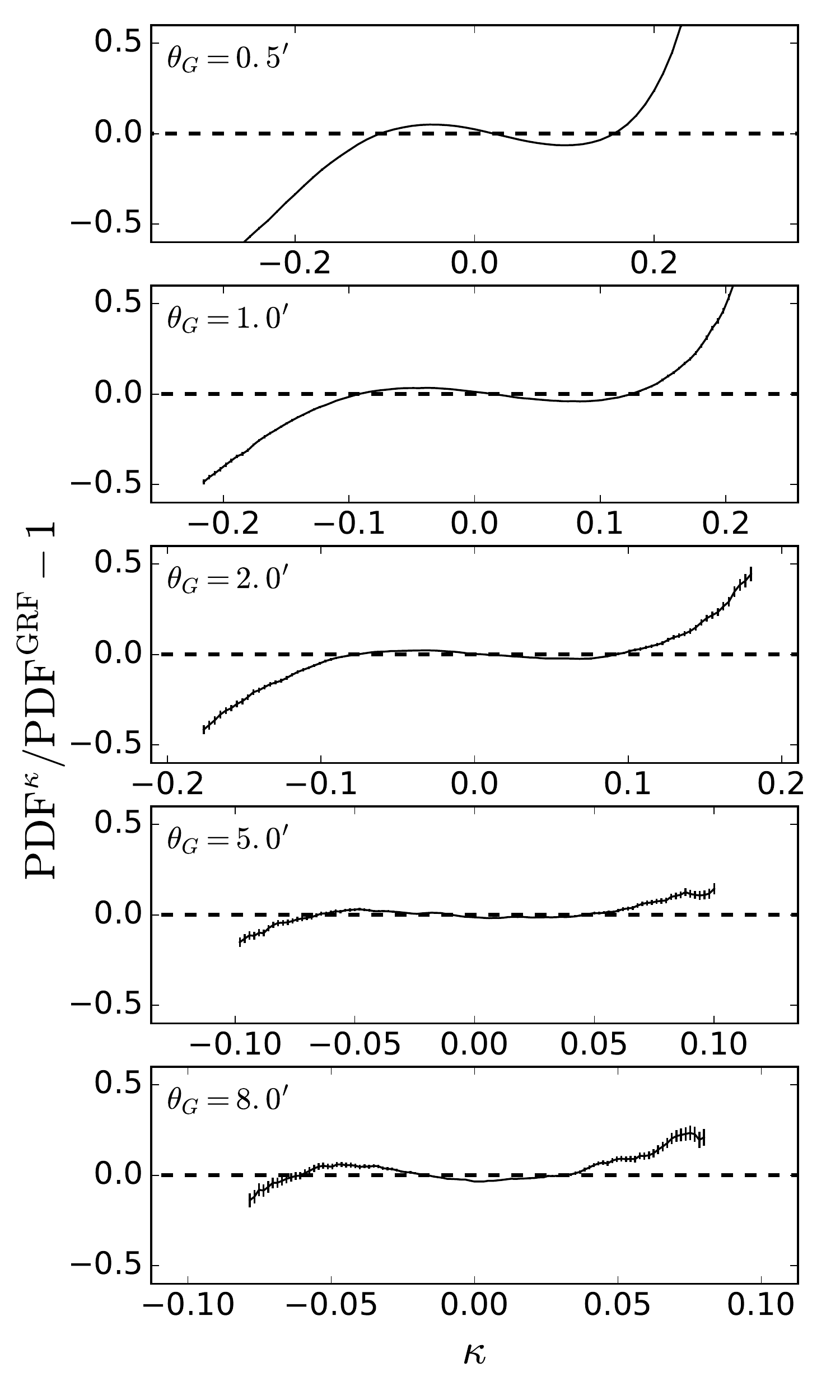}
\end{center}
\caption{\label{fig:noiseless_PDF} PDFs (left panels) of the $N$-body-derived convergence maps and of the Gaussian random fields (labeled ``GRF'') for the noiseless fiducial model, for various smoothing scales (FWHM = 0.5--8 arcmin, top to bottom). Their fractional difference is shown in the right panels. The error bars are scaled to AdvACT sky coverage (20,000 deg$^2$), and are only shown in the right panels for clarity.  Note that no noise is present here, and thus the error bars correspond to a sample-variance-limited survey covering roughly half the sky.}
\end{figure*}

\begin{figure*}
\begin{center}
\includegraphics[width=0.48\textwidth]{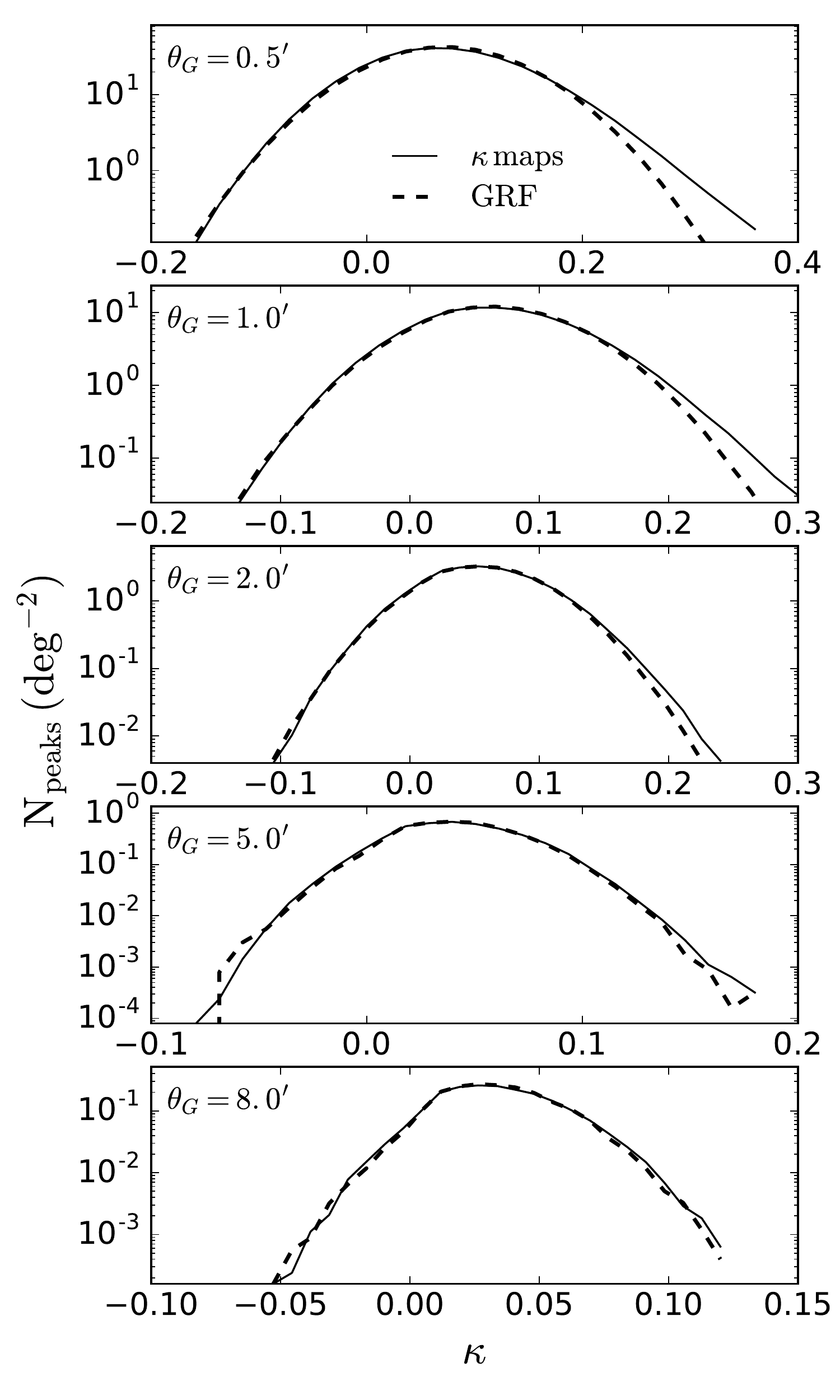}
\includegraphics[width=0.48\textwidth]{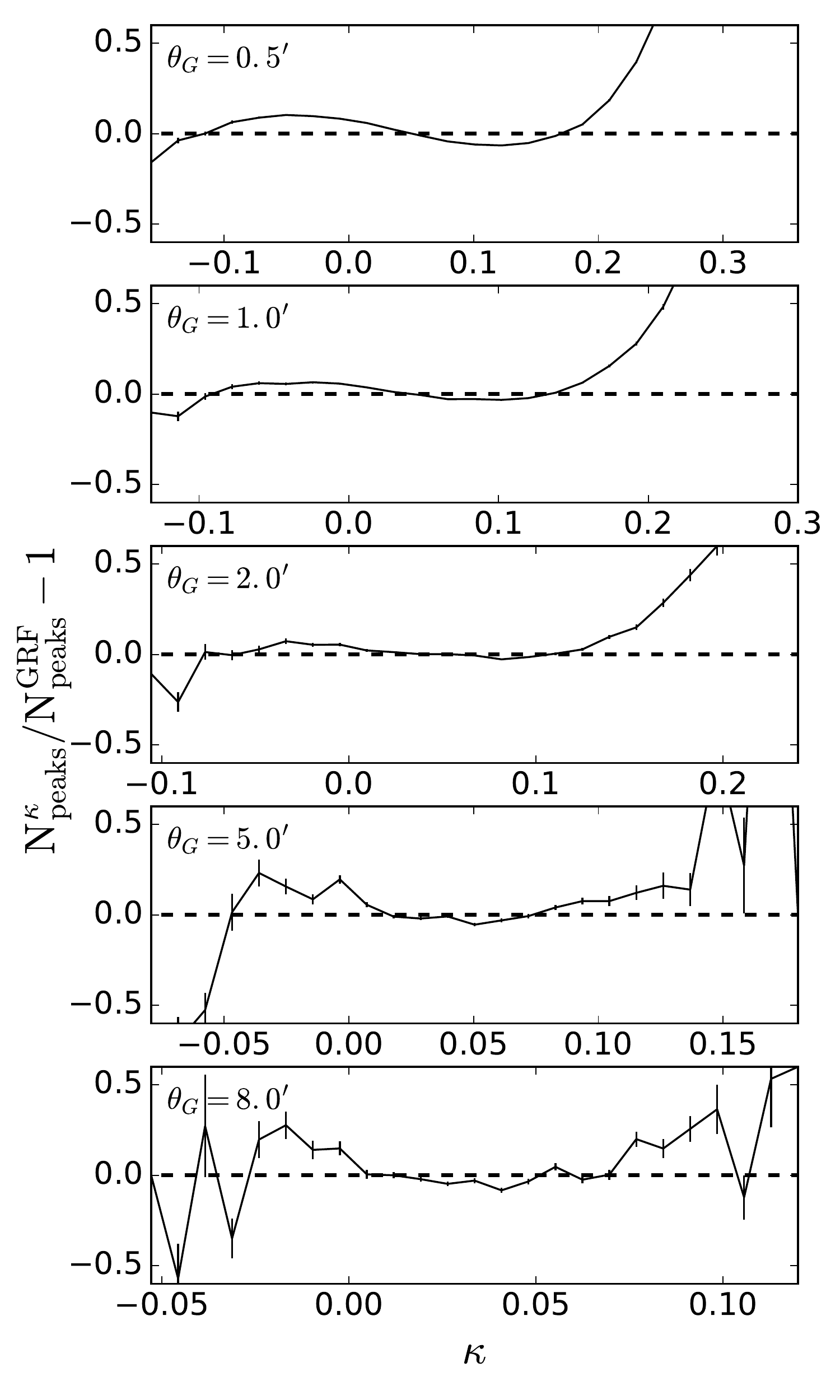}
\end{center}
\caption{\label{fig:noiseless_pk} Same as Fig.~\ref{fig:noiseless_PDF} but for peak counts.}
\end{figure*}

We show the PDF of noiseless $N$-body $\kappa$ maps (PDF$^\kappa$) for the fiducial cosmology in Fig.~\ref{fig:noiseless_PDF}, as well as that of GRF $\kappa$ maps (PDF$^{\rm GRF}$) generated from a power spectrum matching that of the $N$-body-derived maps. To better demonstrate the level of non-Gaussianity, we also show the fractional difference of PDF$^\kappa$ from PDF$^{\rm GRF}$. The error bars are scaled to AdvACT sky coverage (20,000 deg$^2$), though note that no noise is present here.

The departure of PDF$^\kappa$ from the Gaussian case is significant for all smoothing scales examined (FWHM = 0.5--8.0 arcmin), with increasing significance towards smaller smoothing scales, as expected. The excess in high $\kappa$ bins is expected as the result of nonlinear gravitational evolution, echoed by the deficit in low $\kappa$ bins.

We show the comparison of the peak counts of $N$-body $\kappa$ maps (${\rm N}^\kappa_{\rm peaks}$) versus that of GRFs (${\rm N}^{\rm GRF}_{\rm peaks}$) in Fig.~\ref{fig:noiseless_pk}. The difference between ${\rm N}^\kappa_{\rm peaks}$ and ${\rm N}^{\rm GRF}_{\rm peaks}$ is less significant than the PDF, because the number of peaks is much smaller than the number of pixels --- hence, the peak counts have larger Poisson noise. A similar trend of excess (deficit) of high (low) peaks is also seen in $\kappa$ peaks, when compared to the GRF peaks.

\subsection{Covariance matrix}\label{sec:covariance}

\begin{figure}
\begin{center}
\includegraphics[width=0.48\textwidth]{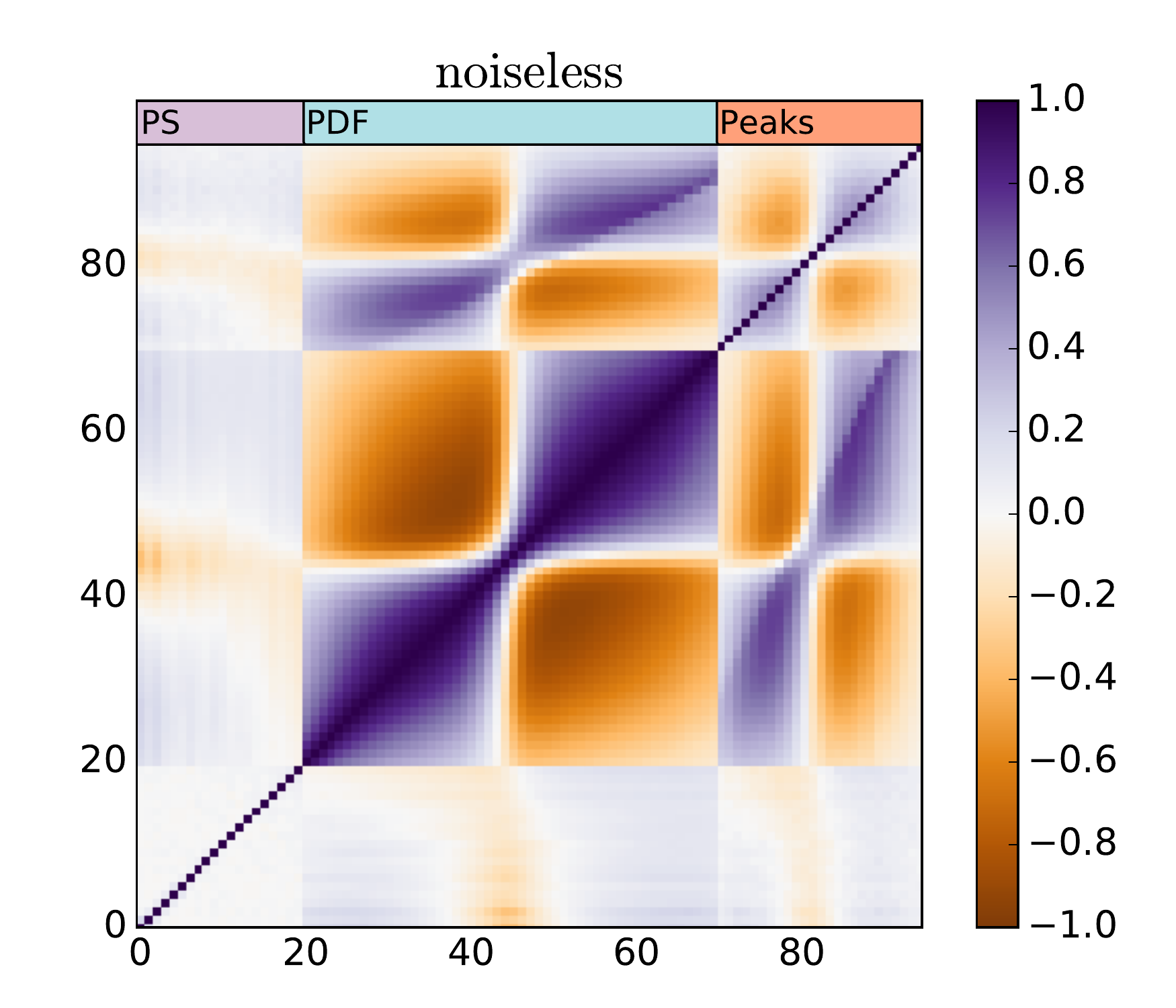}
\includegraphics[width=0.48\textwidth]{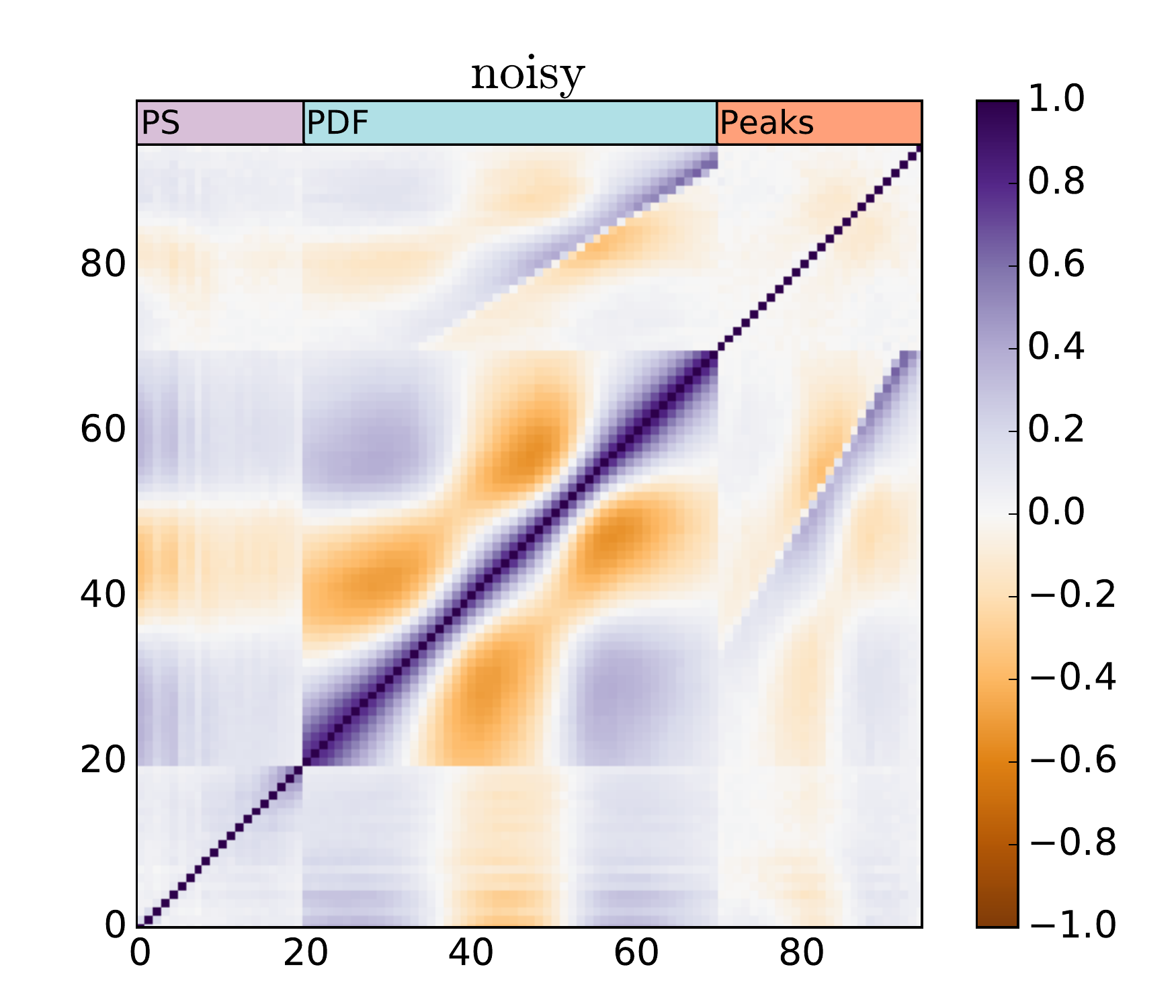}
\end{center}
\caption{\label{fig:corr_mat} Correlation coefficients determined from the full noiseless (top) and noisy (bottom) covariance matrices. Bins 1-20 are for the power spectrum (labeled ``PS''); bins 21-70 are for the PDF; and bins 71-95 are for peak counts.}
\end{figure}

Fig.~\ref{fig:corr_mat} shows the correlation coefficients of the total covariance matrix for both the noiseless and noisy maps,
\begin{align}
\rhoB_{ij} = \frac{\CB_{ij}}{\sqrt{\CB_{ii}\CB_{jj}}}
\end{align}
where $i$ and $j$ denote the bin number, with the first 20 bins for the power spectrum, the next 50 bins for the PDF, and the last 25 bins for peak counts. 

In the noiseless case, the power spectrum shows little covariance in both its own off-diagonal terms ($<10\%$) and cross-covariance with the PDF and peaks ($<20\%$), hinting that the PDF and peaks contain independent information that is beyond the power spectrum. In contrast, the PDF and peak statistics show higher correlation in both self-covariance (i.e., the covariance within the sub-matrix for that statistic only) and cross-covariance, with strength almost comparable to the diagonal components. They both show strong correlation between nearby $\kappa$ bins (especially in the moderate-$|\kappa|$ regions), which arises from contributions due to common structures amongst the bins (e.g., galaxy clusters).  Both statistics show anti-correlation between positive and negative $\kappa$ bins. The anti-correlation may be due to mass conservation --- e.g., large amounts of mass falling into halos would result in large voids in surrounding regions.  

In the noisy case, the off-diagonal terms are generally smaller than in the noiseless case.  Moreover, the anti-correlation seen previously between the far positive and negative $\kappa$ tails in the PDF is now a weak positive correlation --- we attribute this difference to the complex non-Gaussianity of the reconstruction noise.  Interestingly, the self-covariance of the peak counts is significantly reduced compared to the noiseless case, while the self-covariance of the PDF persists to a reasonable degree.

\subsection{Effect of reconstruction noise}\label{sec:recon_noise}

\begin{figure}
\begin{center}
\includegraphics[width=0.48\textwidth]{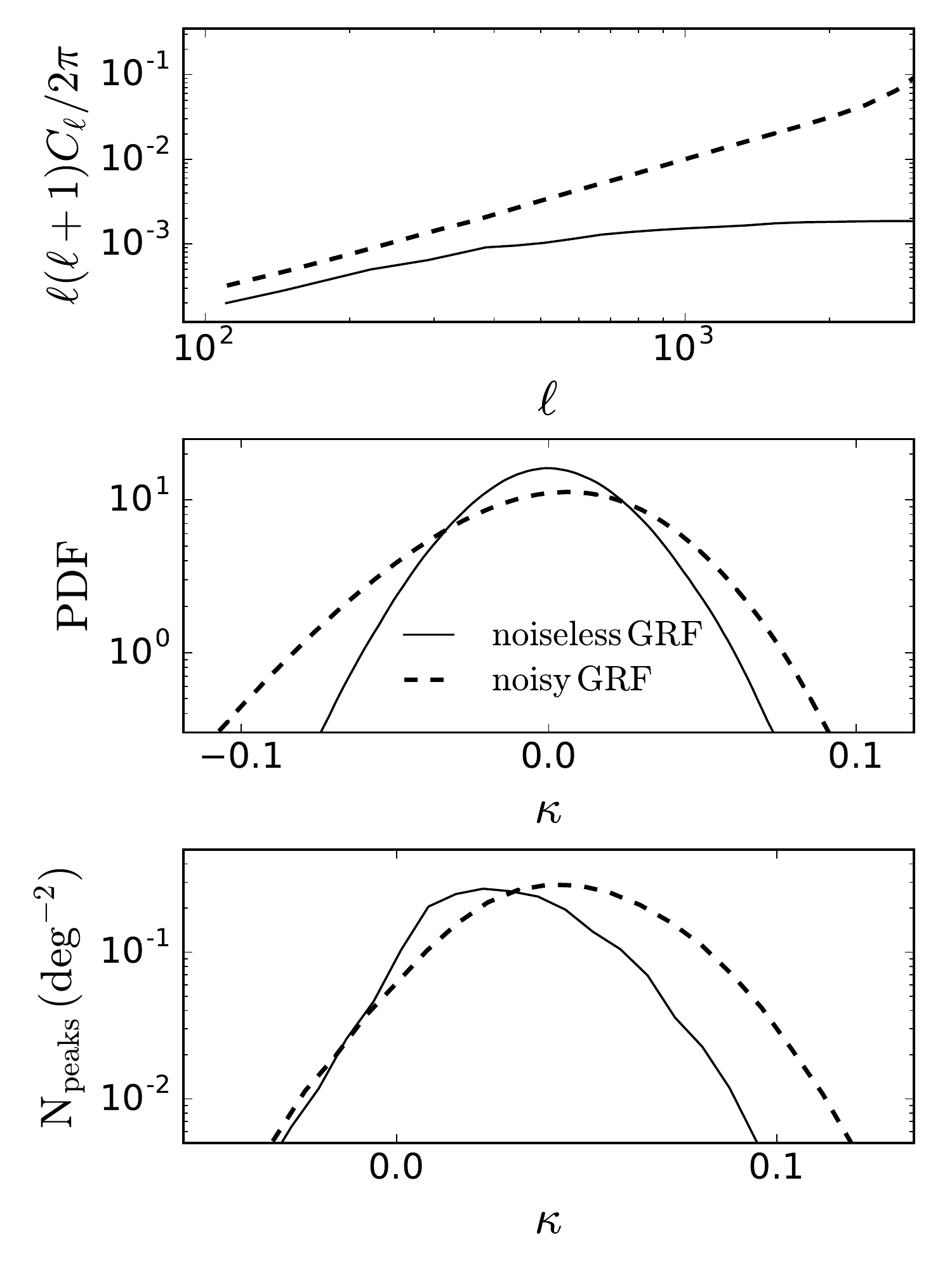}
\end{center}
\caption{\label{fig:recon} We demonstrate the effect of reconstruction noise on the power spectrum (top), the PDF (middle), and peak counts (bottom) by using Gaussian random field $\kappa$ maps (rather than $N$-body-derived maps) as input to the reconstruction pipeline. The noiseless (solid curves) and noisy/reconstructed (dashed curves) statistics are shown.  All maps used here have been smoothed with a Gaussian kernel of FWHM $= 8$ arcmin.} 
\end{figure}

To disentangle the effect of reconstruction noise from that of nonlinear structure growth, we compare the three statistics before (noiseless) and after (noisy) reconstruction, using only the GRF $\kappa$ fields. Fig.~\ref{fig:recon} shows the power spectra, PDFs, and peak counts for both the noiseless (solid curves) and noisy (dashed curves) GRFs, all smoothed with a FWHM $= 8$ arcmin Gaussian window. The reconstructed power spectrum has significant noise on small scales, as expected (this is dominated by the usual ``$N^{(0)}$'' noise bias).

The post-reconstruction PDF shows skewness, defined as 
\begin{equation}
\label{eq.skewdef}
S=\left\langle 
\left( \frac {\kappa-\bar{\kappa}}{\sigma_\kappa}\right)^3 \right\rangle,
\end{equation}
which is not present in the input GRFs. In other words, the reconstructed maps have a non-zero three-point function, even though the input GRF $\kappa$ maps in this case do not. While this may seem surprising at first, we recall that the three-point function of the reconstructed map corresponds to a six-point function of the CMB temperature map (in the quadratic estimator formalism).  Even for a Gaussian random field, the six-point function contains non-zero Wick contractions (those that reduce to products of two-point functions).  Propagating such terms into the three-point function of the quadratic estimator for $\kappa$, we find that they do not cancel to zero.  This result is precisely analogous to the usual ``$N^{(0)}$ bias'' on the CMB lensing power spectrum, in which the two-point function of the (Gaussian) primary CMB temperature gives a non-zero contribution to the temperature four-point function.  The result in Fig.~\ref{fig:recon} indicates that the similar PDF ``$N^{(0)}$ bias'' contains a negative skewness (in addition to non-zero kurtosis and higher moments).  While it should be possible to derive this result analytically, we defer the full calculation to future work.  If we filter the reconstructed $\kappa$ maps with a large smoothing kernel, the skewness in the reconstructed PDF is significantly decreased (see Fig.~\ref{fig:skew}).  We briefly investigate the PDF of the Planck 2015 CMB lensing map~\cite{planck2015xv} and do not see clear evidence of such skewness --- we attribute this to the low effective resolution of the Planck map (FWHM $\sim$ few degrees).  Finally, we note that a non-zero three-point function of the reconstruction noise could potentially alter the forecasted $\kappa$ bispectrum results of Ref.~\cite{Namikawa2016} (where the reconstruction noise was taken to be Gaussian).  The non-Gaussian properties of the small-scale reconstruction noise were noted in Ref.~\cite{HuOkamoto2002}, who pointed out that the quadratic estimator at high-$\ell$ is constructed from progressively fewer arcminute-scale CMB fluctuations.

Similarly, the $\kappa$ peak count distribution also displays skewness after reconstruction, although it is less dramatic than that seen in the PDF. The peak of the distribution shifts to a higher $\kappa$ value due to the additional noise in the reconstructed maps. We note that the shape of the peak count distribution becomes somewhat rough when large smoothing kernels are applied to the maps, due to the small number of peaks present in this situation (e.g., $\approx 29$ peaks in a 12.25~deg$^2$ map with FWHM = 8 arcmin Gaussian window).

\subsection{Non-Gaussianity in reconstructed maps}\label{sec:non-gauss_recon}

\begin{figure*}
\begin{center}
\includegraphics[width=0.48\textwidth]{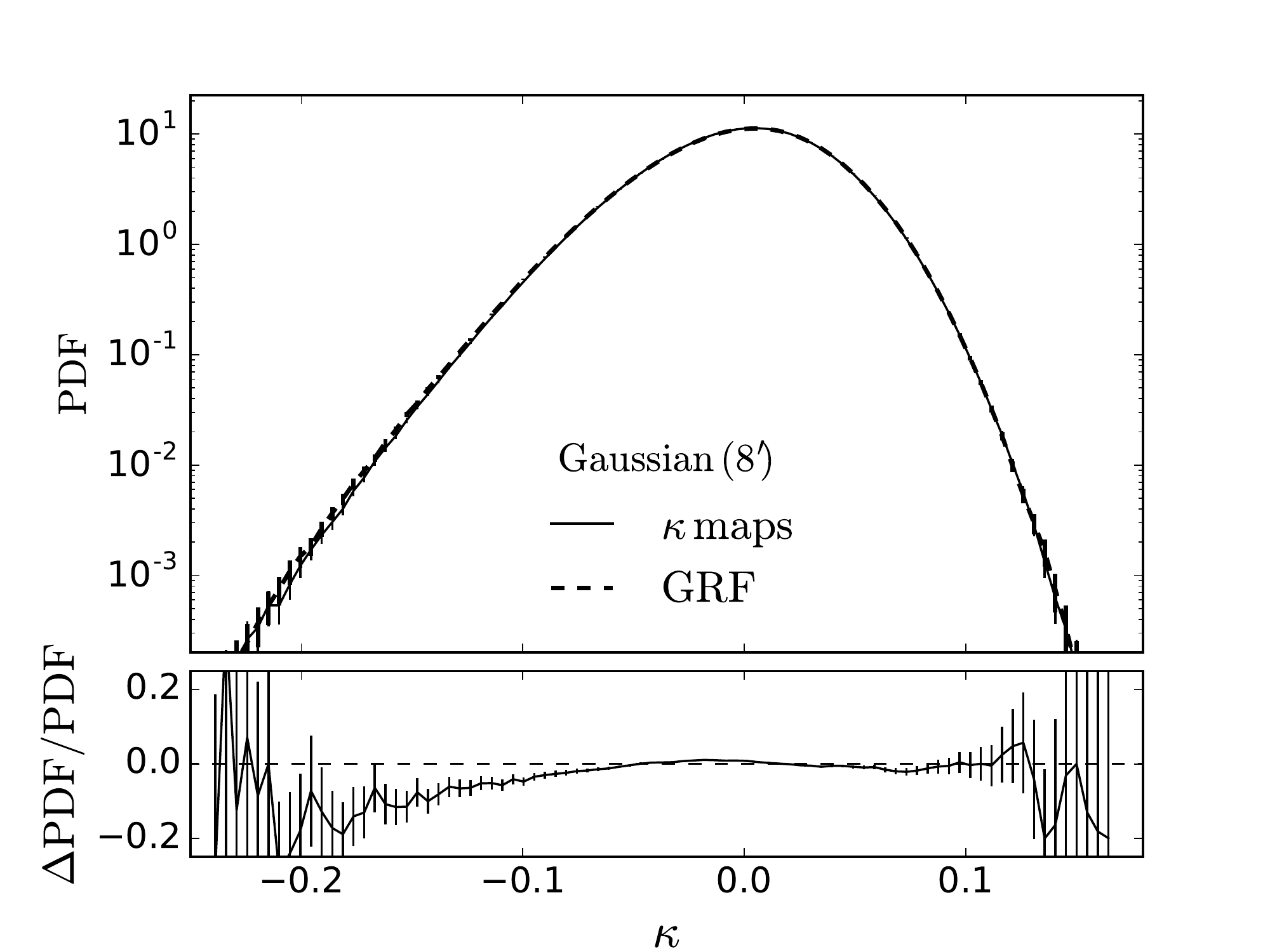}
\includegraphics[width=0.48\textwidth]{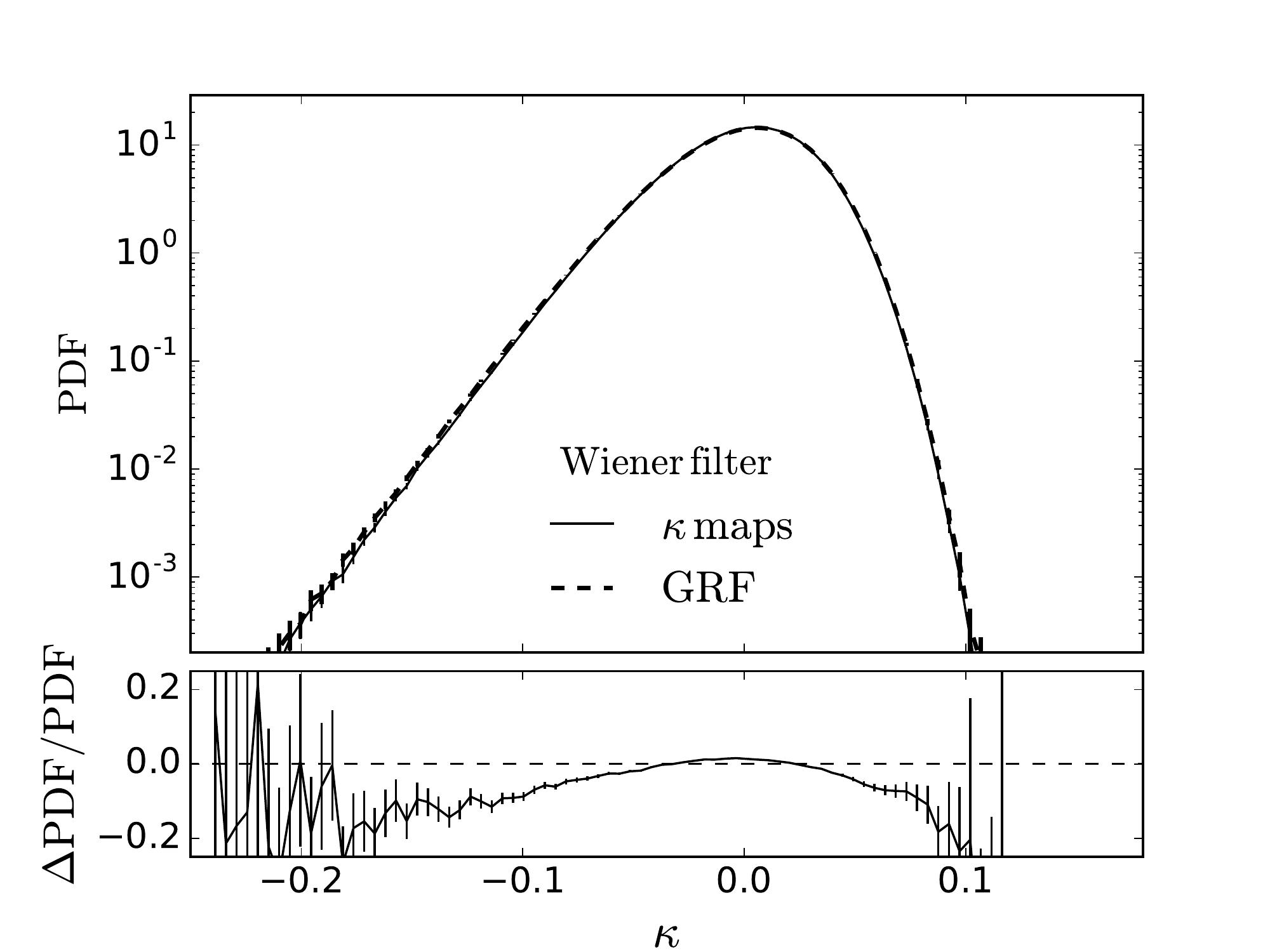}
\end{center}
\caption{\label{fig:noisyPDF} PDFs of the reconstructed convergence maps, when considering input $N$-body-derived $\kappa$ maps (solid curves) or input Gaussian random field $\kappa$ maps (dashed curves).  The difference of the curves is due to nonlinear evolution (and post-Born effects) present in the former maps, but not the latter.  Maps are smoothed with an 8 arcmin Gaussian window (left panel) or by a Wiener filter (right panel). Error bars are scaled to AdvACT sky coverage (20,000 deg$^2$).}
\end{figure*}

\begin{figure*}
\begin{center}
\includegraphics[width=0.48\textwidth]{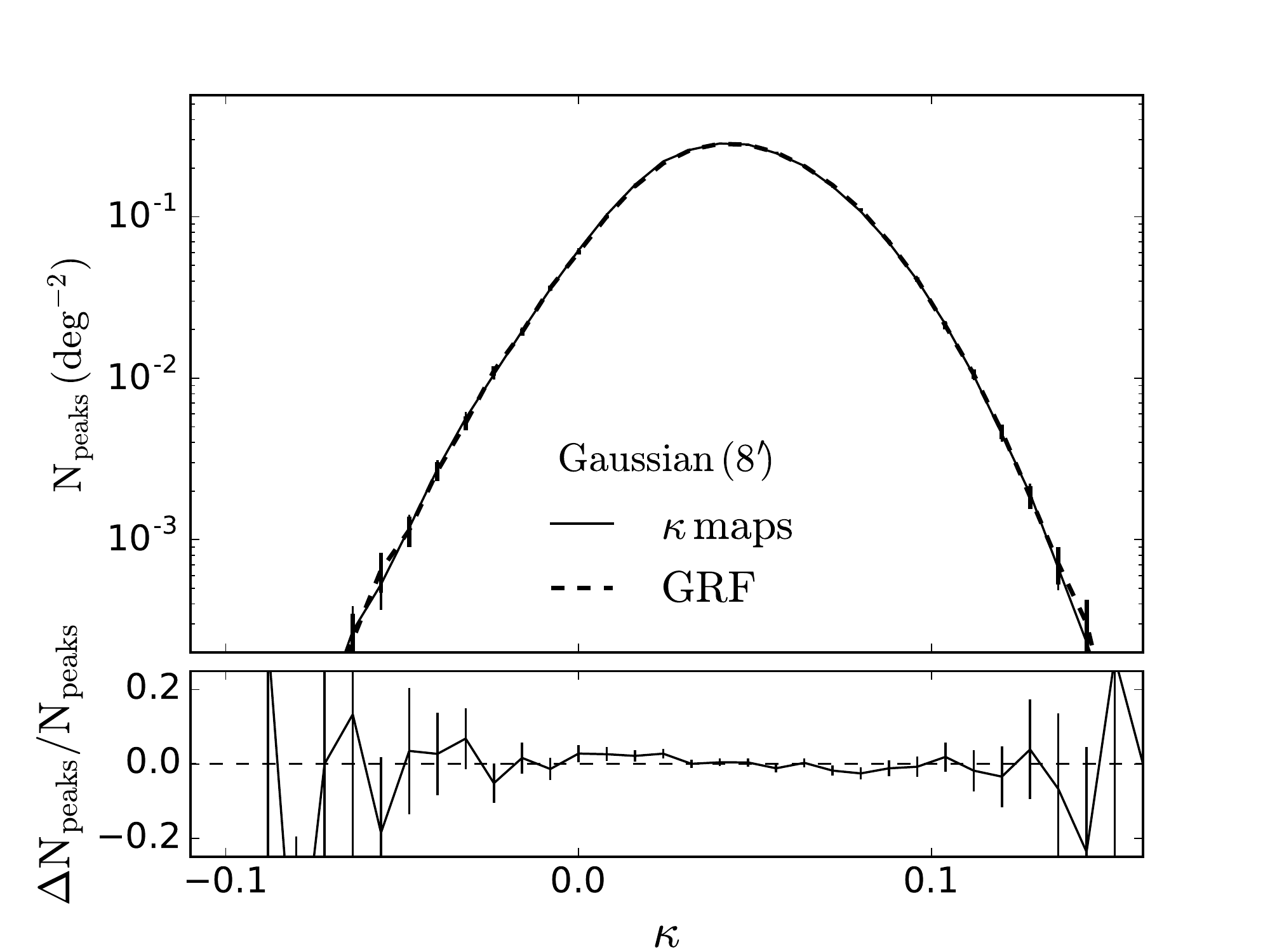}
\includegraphics[width=0.48\textwidth]{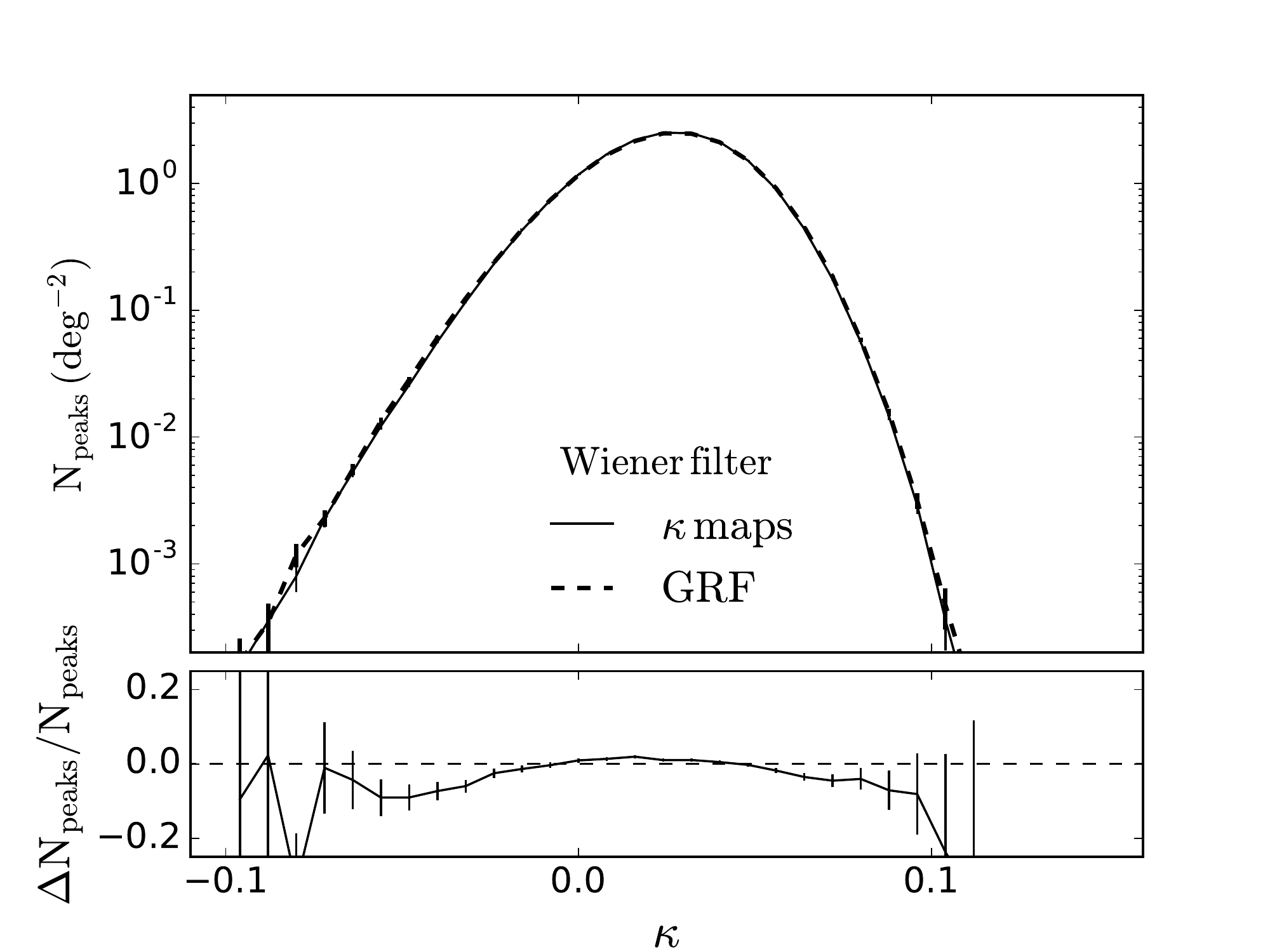}
\end{center}
\caption{\label{fig:noisypk} Same as Fig.~\ref{fig:noisyPDF}, but for peak counts.}
\end{figure*}

We show the PDF and peak counts of the reconstructed $\kappa$ maps in Figs.~\ref{fig:noisyPDF} and \ref{fig:noisypk}, respectively. The left panels of these figures show the results using maps with an 8 arcmin Gaussian smoothing window. We further consider a Wiener filter, which is often used to filter out noise based on some known information in a signal (i.e., the noiseless power spectrum in our case). The right panels show the Wiener-filtered results, where we inverse-variance weight each pixel in Fourier space, i.e., each Fourier mode is weighted by the ratio of the noiseless power spectrum to the noisy power spectrum (c.f. Fig.~\ref{fig:recon}),
\begin{align}
f^{\rm Wiener} (\ell) = \frac{C_\ell^{\rm noiseless}}{C_\ell^{\rm noisy}} \,.
\end{align}

Compared to the noiseless results shown in Figs.~\ref{fig:noiseless_PDF} and~\ref{fig:noiseless_pk}, the differences between the PDF and peaks from the $N$-body-derived $\kappa$ maps and those from the GRF-derived $\kappa$ maps persist, but with less significance. For the Wiener-filtered maps, the deviations of the $N$-body-derived $\kappa$ statistics from the GRF case are 9$\sigma$ (PDF) and 6$\sigma$ (peaks), where we derived the significances using the simulated covariance from the $N$-body maps
\footnote
{We note that the signal-to-noise ratios predicted here are comparable to the $\approx 7\sigma$ bispectrum prediction that would be obtained by rescaling the SPT-3G result from Table I of Ref.~\cite{Pratten2016} to the AdvACT sky coverage (which is a slight overestimate given AdvACT's higher noise level).  The higher significance for the PDF found here could be due to several reasons: (i) additional contributions to the signal-to-noise for the PDF from higher-order polyspectra beyond the bispectrum; (ii) inaccuracy of the nonlinear fitting formula used in Ref.~\cite{Pratten2016} on small scales, as compared to the N-body methods used here; (iii) reduced cancellation between the nonlinear growth and post-Born effects in higher-order polyspectra (for the bispectrum, these contributions cancel to a large extent, reducing the signal-to-noise~\cite{Pratten2016}).}.  These deviations capture the influence of both nonlinear evolution and post-Born effects.

\begin{figure}
\begin{center}
\includegraphics[width=0.48\textwidth]{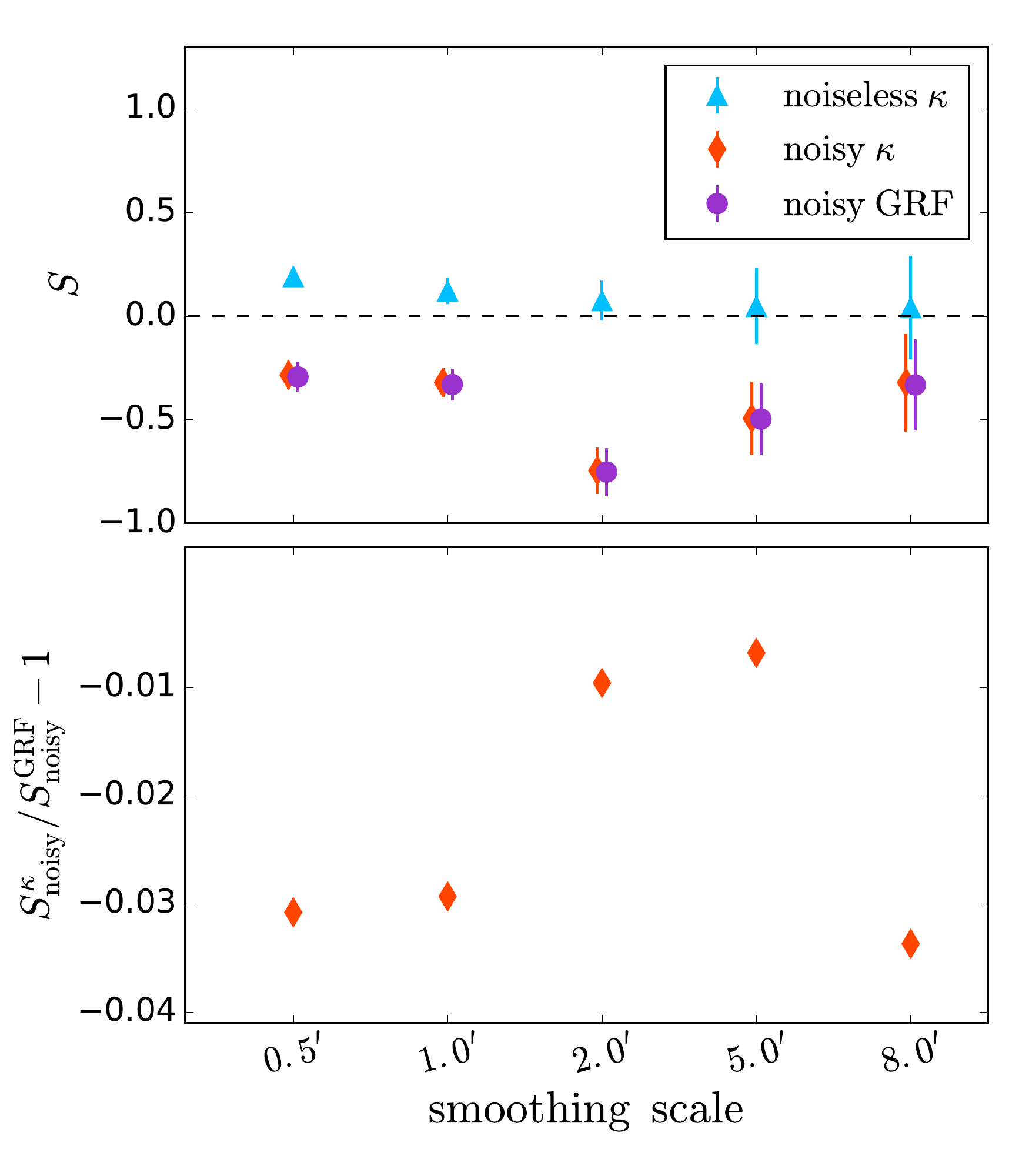}
\end{center}
\caption{\label{fig:skew} Top panel: the skewness of the noiseless (triangles) and reconstructed, noisy (diamonds: $N$-body $\kappa$ maps; circles: GRF) PDFs. Bottom panel: the fractional difference between the skewness of the reconstructed $N$-body $\kappa$ and the reconstructed GRF.  The error bars are for our map size (12.25 deg$^2$), and are only shown in the top panel for clarity.}
\end{figure}

While the differences between the $N$-body and GRF cases in Figs.~\ref{fig:noisyPDF} and \ref{fig:noisypk} are clear, understanding their detailed structure is more complex.  First, note that the GRF cases exhibit the skewness discussed in Sec.~\ref{sec:recon_noise}, which arises from the reconstruction noise itself.  We show the skewness of the reconstructed PDF (for both the $N$-body and GRF cases) compared with that of the noiseless ($N$-body) PDF for various smoothing scales in Fig.~\ref{fig:skew}.  The noiseless $N$-body maps are positively skewed, as physically expected.  The reconstructed, noisy maps are negatively skewed, for both the $N$-body and GRF cases.  However, the reconstructed $N$-body results are less negatively skewed than the reconstructed GRF results (bottom panel of Fig.~\ref{fig:skew}), presumably because the $N$-body PDF (and peaks) contain contributions from the physical skewness, which is positive (see Figs.~\ref{fig:noiseless_PDF} and~\ref{fig:noiseless_pk}).  However, the physical skewness is not large enough to overcome the negative ``$N^{(0)}$''-type skewness coming from the reconstruction noise.  We attribute the somewhat-outlying point at FWHM $=8$ arcmin in the bottom panel of Fig.~\ref{fig:skew} to a noise fluctuation, as the number of pixels at this smoothing scale is quite low (the deviation is consistent with zero).  The decrease in $|S|$ between the FWHM $=2$ arcmin and 1 arcmin cases in the top panel of Fig.~\ref{fig:skew} for the noisy maps is due to the large increase in $\sigma_{\kappa}$ between these smoothing scales, as the noise is blowing up on small scales.  The denominator of Eq.~(\ref{eq.skewdef}) thus increases dramatically, compared to the numerator.

Comparisons between the reconstructed PDF in the $N$-body case and GRF case are further complicated by the fact that higher-order ``biases'' arise due to the reconstruction.  For example, the skewness of the reconstructed $N$-body $\kappa$ receives contributions from many other terms besides the physical skewness and the ``$N^{(0)}$ bias'' described above --- there will also be Wick contractions involving combinations of two- and four-point functions of the CMB temperature and $\kappa$ (and perhaps an additional bias coming from a different contraction of the three-point function of $\kappa$, analogous to the ``$N^{(1)}$'' bias for the power spectrum~\cite{Hanson2011}). So the overall ``bias'' on the reconstructed skewness will differ from that in the simple GRF case.  This likely explains why we do not see an excess of positive $\kappa$ values over the GRF case in the PDFs shown in Fig.~\ref{fig:noisyPDF}.  While this excess is clearly present in the noiseless case (Fig.~\ref{fig:noiseless_PDF}), and it matches physical intuition there, the picture in the reconstructed case is not simple, because there is no guarantee that the reconstruction biases in the $N$-body and GRF cases are exactly the same.  Thus, a comparison of the reconstructed $N$-body and GRF PDFs contains a mixture of the difference in the biases and the physical difference that we expect to see.  Similar statements hold for comparisons of the peak counts.

Clearly, a full accounting of all such individual biases would be quite involved, but the key point here is that all these effects are fully present in our end-to-end simulation pipeline.  While an analytic understanding would be helpful, it is not necessary for the forecasts we present below.

\subsection{Cosmological constraints}\label{sec:constraints}

Before we proceed to present the cosmological constraints from non-Gaussian statistics, it is necessary to do a sanity check by comparing the forecasted contour from our simulated power spectra to that from an analytic Fisher estimate,
\begin{align}
\FB_{\alpha \beta}=\frac{1}{2} {\rm Tr}
\left\{\CB^{-1}_{\rm Gauss} 
\left[\left(\frac {\partial C_\ell}{\partial p_\alpha} \right)
\left(\frac {\partial C_\ell}{\partial p_\beta}\right)^T+ 
\left(\alpha\leftrightarrow\beta \right)
\right]\right\},
\end{align}
where $\left\{ \alpha,\beta \right\} = \left\{ \Omega_m,\sigma_8 \right\}$ and the trace is over $\ell$ bins. $\CB_{\rm Gauss}$ is the Gaussian covariance matrix, with off-diagonal terms set to zero, and diagonal terms equal to the Gaussian variance, 
\begin{align}
\sigma^2_\ell=\frac{2(C_\ell+N_\ell)^2}{f_{\rm sky}(2\ell+1)\Delta\ell}
\end{align}

\begin{figure}
\begin{center}
\includegraphics[width=0.48\textwidth]{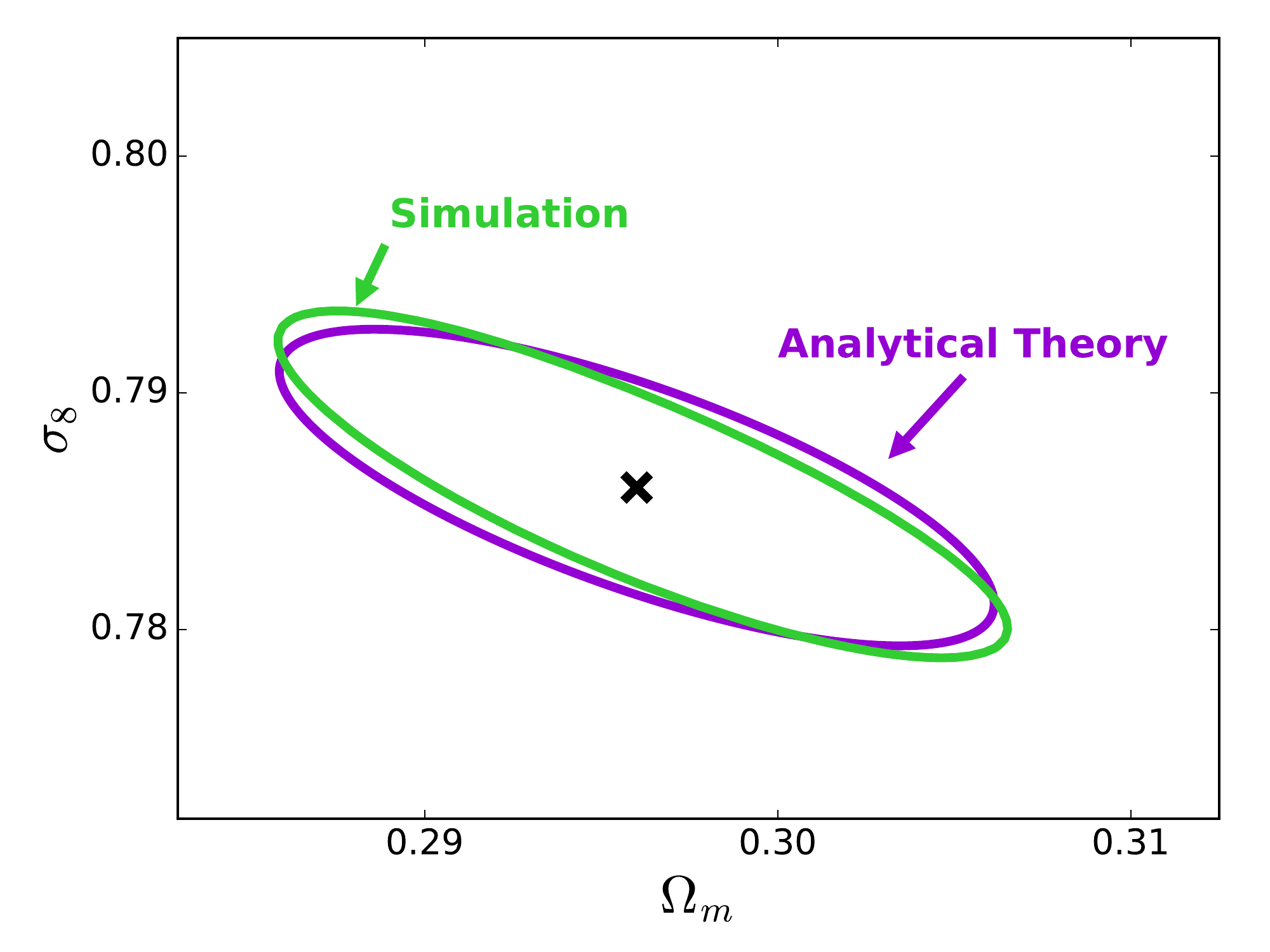}
\end{center}
\caption{\label{fig:contour_fisher} 68\% C.L. contours from an AdvACT-like CMB lensing power spectrum measurement.  The excellent agreement between the simulated and analytic results confirms that non-Gaussian covariances arising from nonlinear growth and reconstruction noise do not strongly bias current analytic CMB lensing power spectrum forecasts (up to $\ell = 2,000$).}
\end{figure}

We compute the theoretical power spectrum $C_\ell$ using the HaloFit model~\cite{Smith2003,Takahashi2012}, with fractional parameter variations of $+1$\% to numerically obtain $\partial C_\ell / \partial p$. $N_\ell$ is the reconstruction noise power spectrum, originating from primordial CMB fluctuations and instrumental/atmospheric noise (note that we only consider white noise here). The sky fraction $f_{\rm sky}=0.485$ corresponds to the 20,000 deg$^2$ coverage expected for AdvACT. $(F^{-1}_{\alpha\alpha})^{\frac{1}{2}}$ is the marginalized error on parameter $\alpha$. Both theoretical and simulated contours use the power spectrum within the $\ell$ range of [100, 2,000]. The comparison is shown in Fig.~\ref{fig:contour_fisher}. The contour from full $N$-body simulations shows good agreement with the analytical Fisher contour.  This result indicates that approximations made in current analytical CMB lensing power spectrum forecasts are accurate, in particular the neglect of non-Gaussian covariances from nonlinear growth.  A comparison of the analytic and reconstructed power spectra will be presented in Ref.~\cite{Sherwin2016}.

\begin{figure*}
\begin{center}
\includegraphics[width=0.48\textwidth]{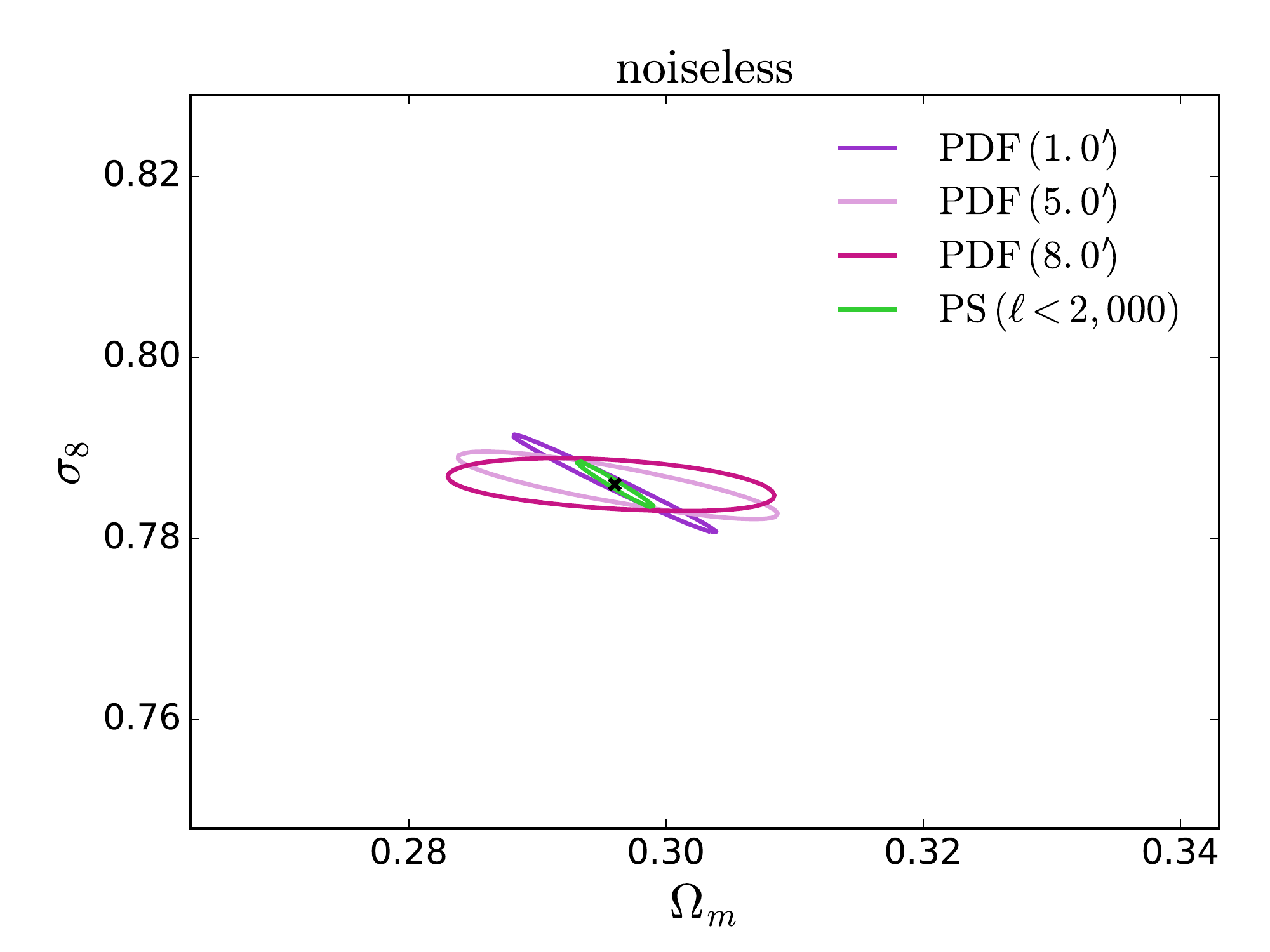}
\includegraphics[width=0.48\textwidth]{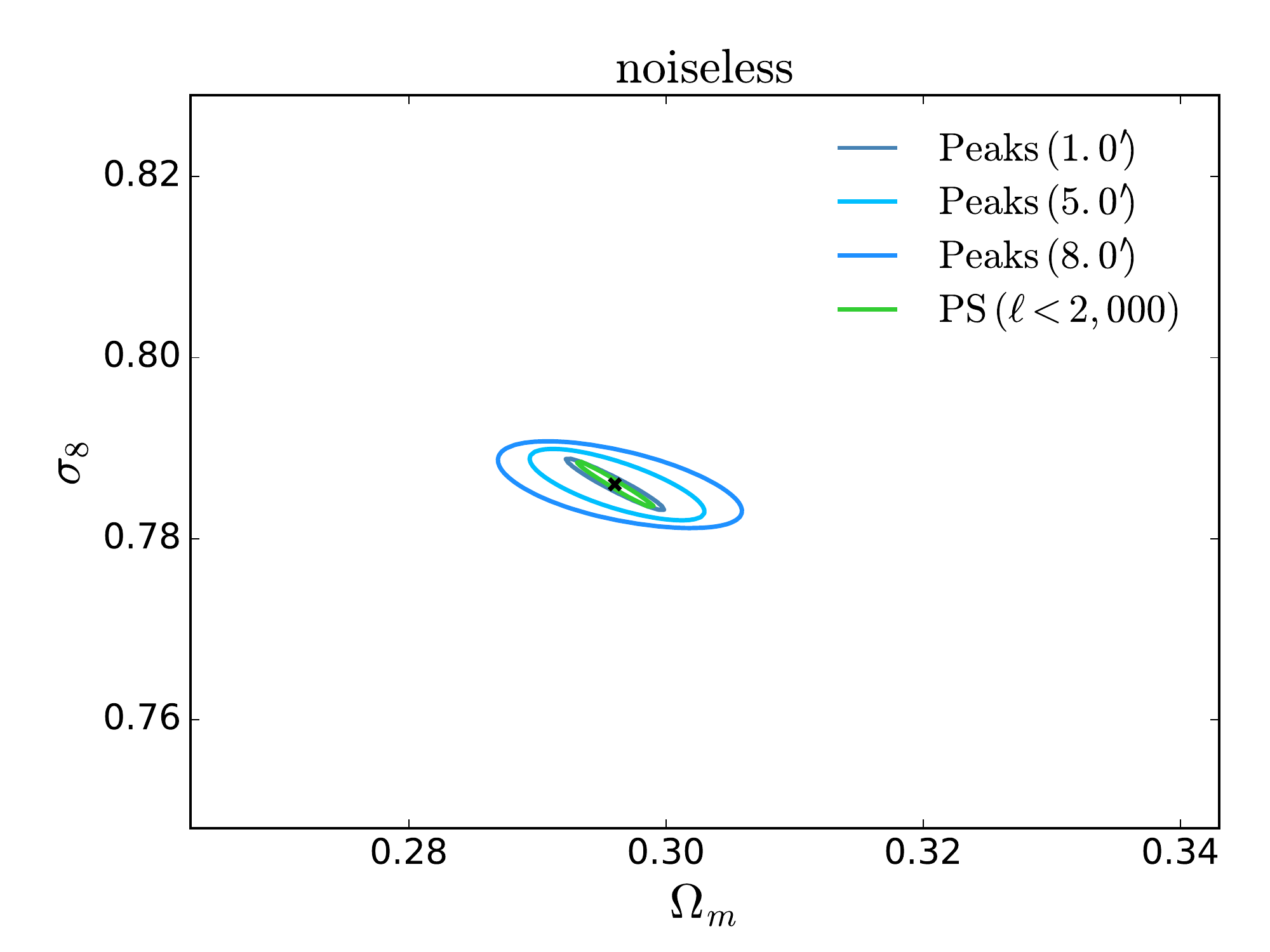}
\end{center}
\caption{\label{fig:contour_noiseless} 68\% C.L. contours derived from the noiseless PDF (left) and peak counts (right) using three smoothing scales, compared with that from the noiseless power spectrum ($\ell<2,000$). Three Gaussian smoothing scales are tested (FWHM = 1.0, 5.0, 8.0 arcmin). The contours are scaled to AdvACT sky coverage of 20,000 deg$^2$. Note that we use the same scale as in Fig.~\ref{fig:contour_noisy} (where we show the contours derived from the reconstructed noisy maps), in order to demonstrate the contour size change due to the noise.}
\end{figure*}

Fig.~\ref{fig:contour_noiseless} shows contours derived using noiseless maps for the PDF and peak count statistics, compared with that from the noiseless power spectrum. We compare three different smoothing scales (1.0, 5.0, 8.0 arcmin), and find that smaller smoothing scales have stronger constraining power. However, even with the smallest smoothing scale (1.0 arcmin), the PDF contour is still significantly larger than that of the power spectrum. Peak counts using 1.0 arcmin smoothing show almost equivalent constraining power as the power spectrum. However, we note that 1.0 arcmin smoothing is not a fair comparison to the power spectrum with cutoff at $\ell<2,000$, because in reality, the beam size and instrument noise is likely to smear out signals smaller than a few arcmin scale (see below).

\begin{figure}
\begin{center}
\includegraphics[width=0.48\textwidth]{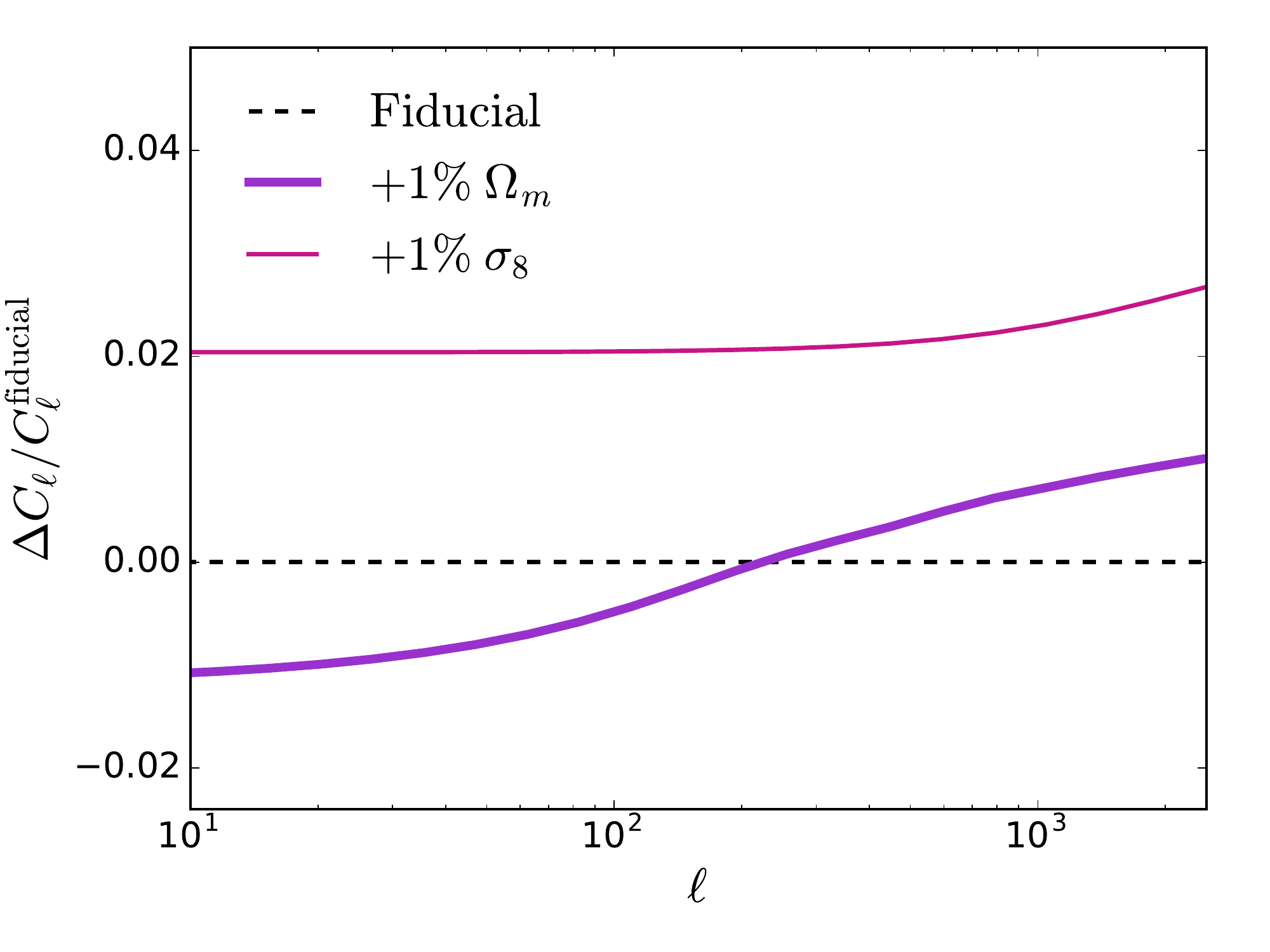}
\end{center}
\caption{\label{fig:cell_diff} Fractional difference of the CMB lensing power spectrum after a 1\% increase in $\Omega_m$ (thick solid line) or $\sigma_8$ (thin solid line), compared to the fiducial power spectrum. Other parameters are fixed at their fiducial values.}
\end{figure}

At first, it may seem surprising that the PDF is not at least as constraining as the power spectrum in Fig.~\ref{fig:contour_noiseless}, since the PDF contains the information in the variance.  However, this only captures an overall amplitude of the two-point function, whereas the power spectrum contains scale-dependent information.\footnote{Note that measuring the PDF or peak counts for different smoothing scales can recover additional scale-dependent information as well.}  We illustrate this in Fig.~\ref{fig:cell_diff}, where we compare the fiducial power spectrum to that with a 1\% increase in $\Omega_m$ or $\sigma_8$ (while keeping other parameters fixed).  While $\sigma_8$ essentially re-scales the power spectrum by a factor $\sigma_8^2$, apart from a steeper dependence at high-$\ell$ due to nonlinear growth, $\Omega_m$ has a strong shape dependence.  This is related to the change in the scale of matter-radiation equality~\cite{planck2015xv}.  Thus, for a noiseless measurement, the shape of the power spectrum contains significant additional information about these parameters, which is not captured by a simple change in the overall amplitude of the two-point function.  This is the primary reason that the power spectrum is much more constraining than the PDF in Fig.~\ref{fig:contour_noiseless}.

\begin{figure*}
\begin{center}
\includegraphics[width=0.48\textwidth]{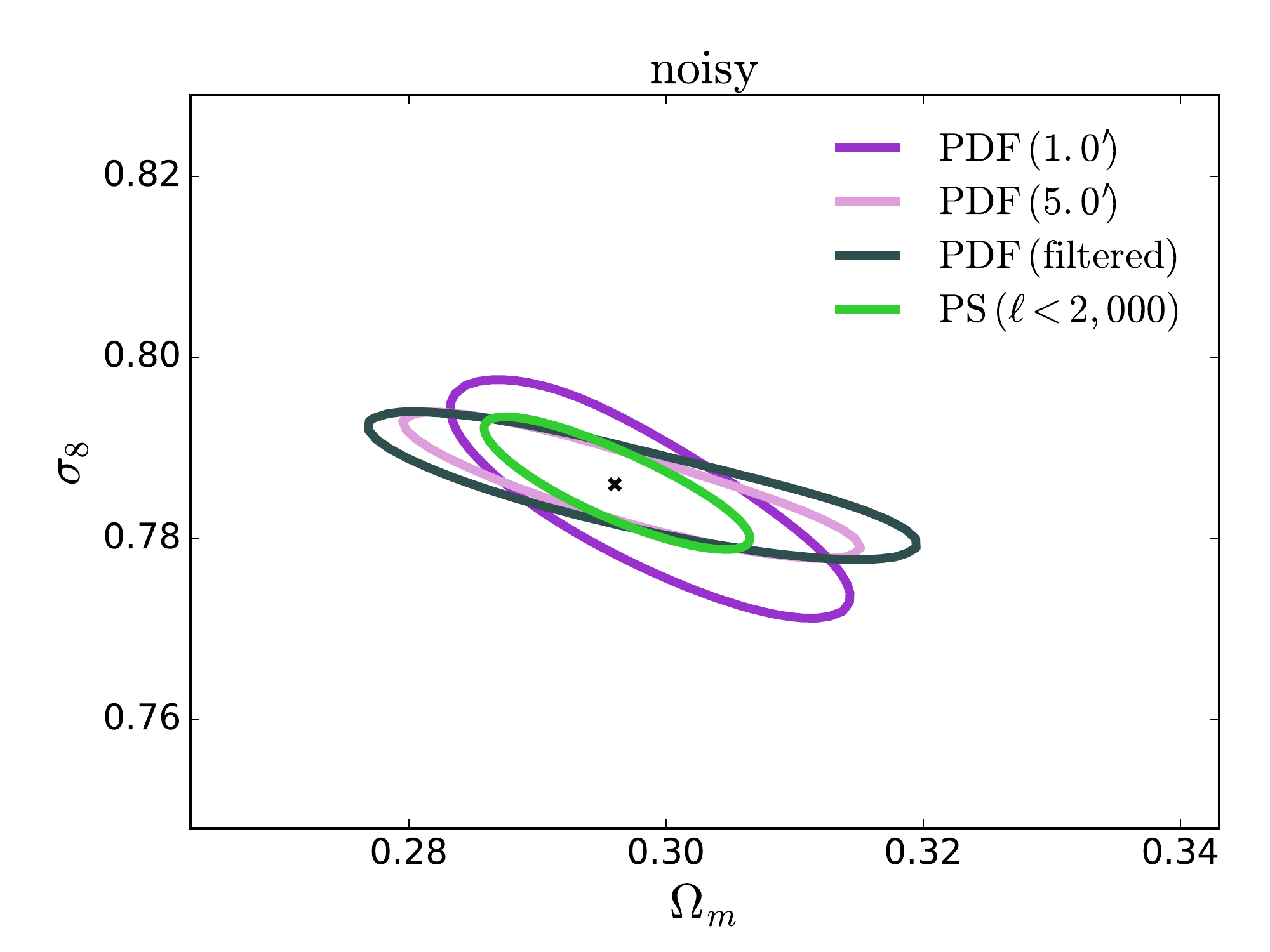}
\includegraphics[width=0.48\textwidth]{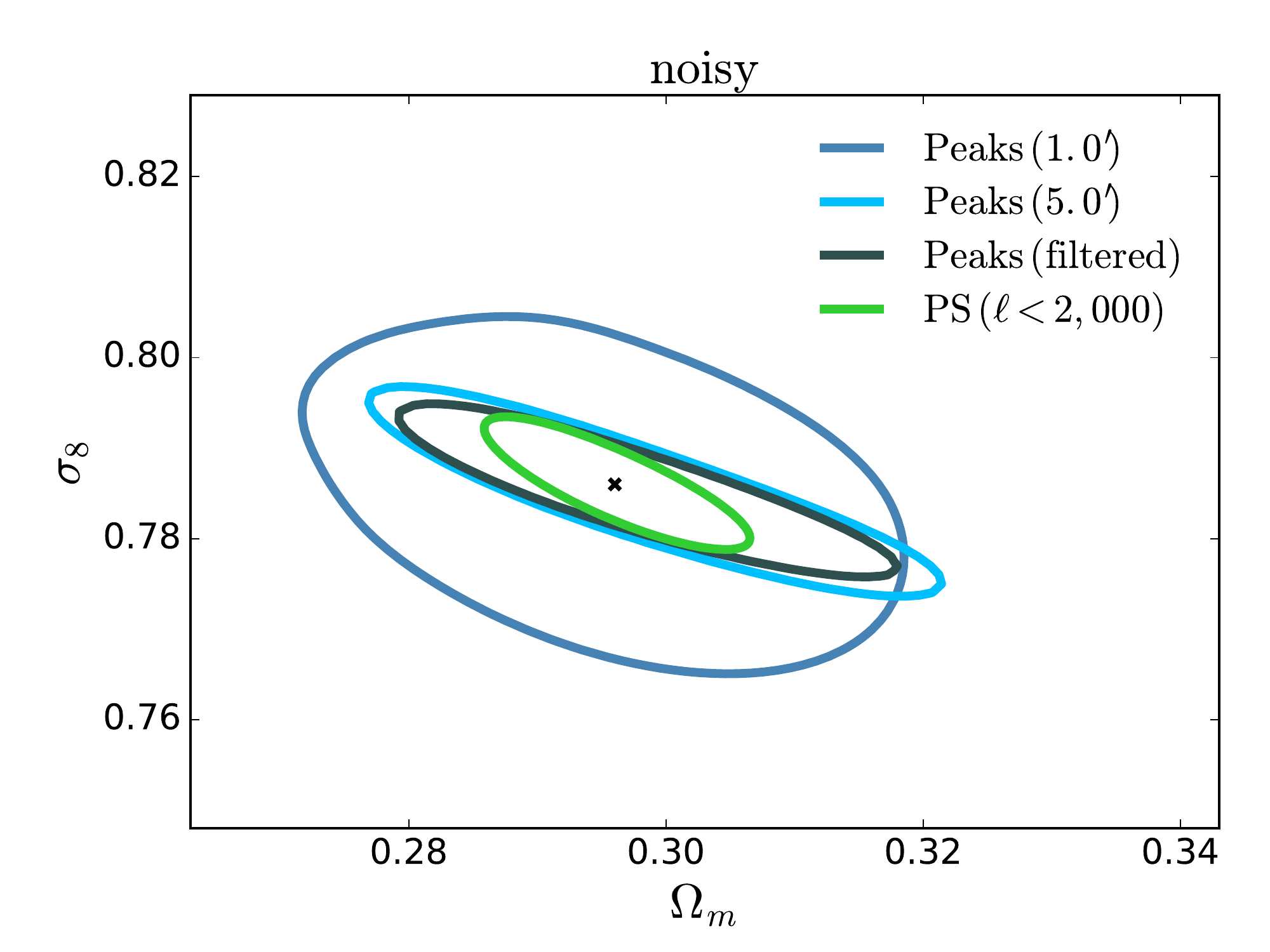}
\end{center}
\caption{\label{fig:contour_noisy} 68\% C.L. contours derived from the reconstructed (noisy) PDF (left) and peak counts (right), compared with that from the power spectrum ($\ell<2,000$). Three map filtering schemes are tested --- Gaussian smoothing with FWHM of 1.0 and 5.0 arcmin  and the Wiener filter. The contours are scaled to AdvACT sky coverage of 20,000 deg$^2$.  Note that we use the same scale as in Fig.~\ref{fig:contour_noiseless} (where we show the contours derived from the noiseless maps), in order to demonstrate the contour size change due to the noise.}
\end{figure*}

\begin{figure}
\begin{center}
\includegraphics[width=0.48\textwidth]{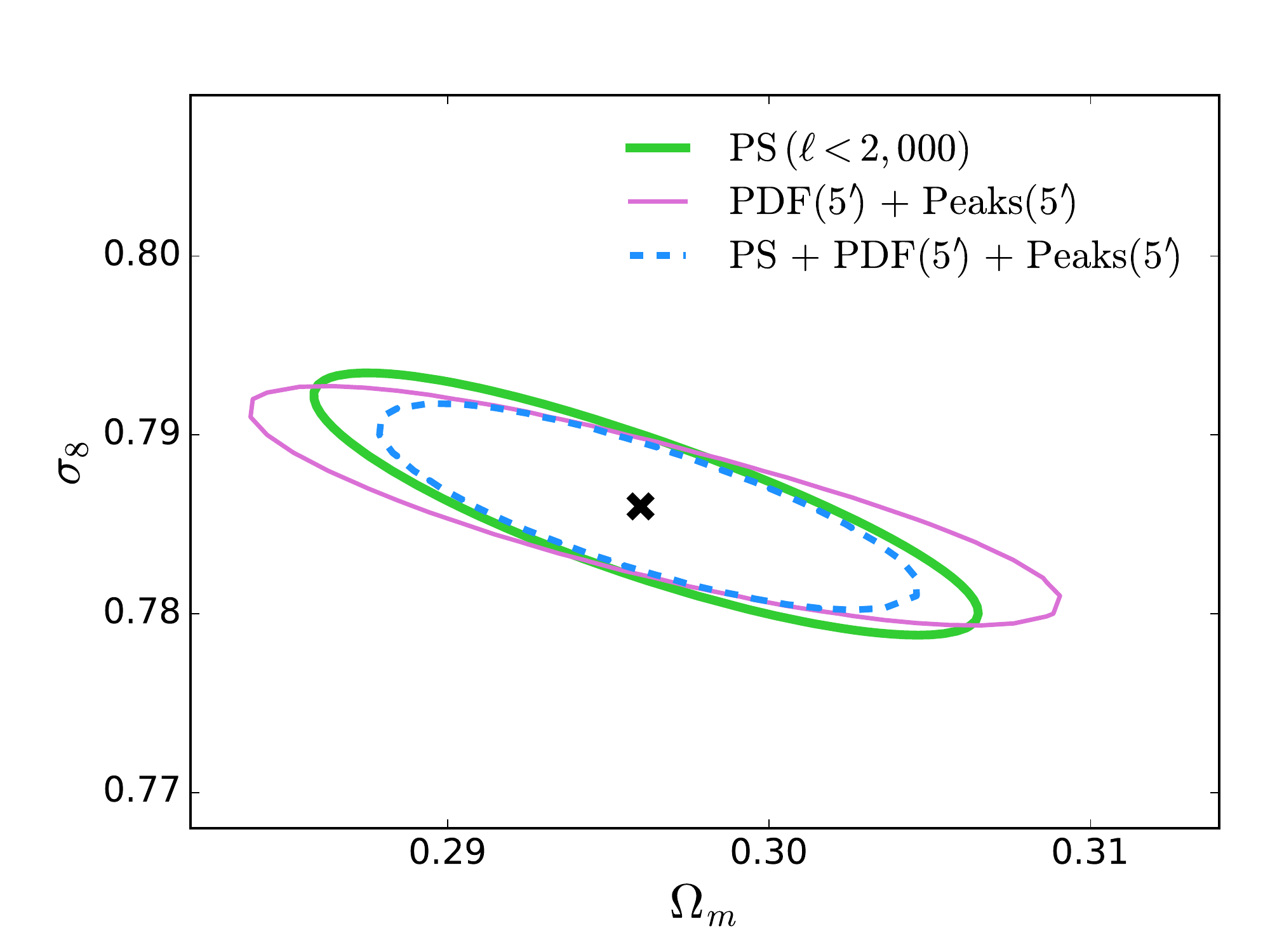}
\end{center}
\caption{\label{fig:contour_comb} 68\% C.L. contours derived using two combinations of the power spectrum, PDF, and peak counts, compared to using the power spectrum alone. Reconstruction noise corresponding to an AdvACT-like survey is included. The contours are scaled to AdvACT sky coverage of 20,000 deg$^2$.}
\end{figure}

Fig.~\ref{fig:contour_noisy} shows contours derived using the reconstructed, noisy $\kappa$ maps. We show results for three different filters --- Gaussian windows of 1.0 and 5.0 arcmin and the Wiener filter. The 1.0 arcmin contour is the worst among all, as noise dominates at this scale. The 5.0 arcmin-smoothed and Wiener-filtered contours show similar constraining power. Using the PDF or peak counts alone, we do not achieve better constraints than using the power spectrum alone, but the parameter degeneracy directions for the statistics are slightly different.  This is likely due to the fact that the PDF and peak counts probe non-linear structure, and thus they have a different dependence on the combination $\sigma_8(\Omega_m)^\gamma$ than the power spectrum does, where $\gamma$ specifies the degeneracy direction.

\begin{table}
\begin{tabular}{|l|c|c|}
\hline
Combination &       $\Delta \Omega_m$  &  $\Delta \sigma_8 $         \\
\hline                                                                                                                  
PS only		&	0.0065	&	0.0044\\
PDF + Peaks	&	0.0076	&      0.0035\\
PS + PDF + Peaks &	0.0045	&      0.0030\\
\hline
\end{tabular}
\caption[]{\label{tab: constraints} Marginalized constraints on $\Omega_m$ and $\sigma_8$ for an AdvACT-like survey from combinations of the power spectrum (PS), PDF, and peak counts, as shown in Fig.~\ref{fig:contour_comb}.}
\end{table}

The error contour derived using all three statistics is shown in Fig.~\ref{fig:contour_comb}, where we use the 5.0 arcmin Gaussian smoothed maps. The one-dimensional marginalized errors are listed in Table~\ref{tab: constraints}.  The combined contour shows moderate improvement ($\approx 30\%$ smaller error contour area) compared to the power spectrum alone.  The improvement is due to the slightly different parameter degeneracy directions for the statistics, which break the $\sigma_8$-$\Omega_m$ degeneracy somewhat more effectively when combined. It is worth noting that we have not included information from external probes that constrain $\Omega_m$ (e.g., baryon acoustic oscillations), which can further break the $\Omega_m$-$\sigma_8$ degeneracy. 

\section{Conclusion}\label{sec:conclude}

In this paper, we use $N$-body ray-tracing simulations to explore the additional information in CMB lensing maps beyond the traditional power spectrum. In particular, we investigate the one-point PDF and peak counts (local maxima in the convergence map). We also apply realistic reconstruction procedures that take into account primordial CMB fluctuations and instrumental noise for an AdvACT-like survey, with sky coverage of 20,000 deg$^2$, noise level 6 $\mu$K-arcmin, and $1.4$ arcmin beam. Our main findings are:

\begin{enumerate}
\item We find significant deviations of the PDF and peak counts of $N$-body-derived $\kappa$ maps from those of Gaussian random field $\kappa$ maps, both in the noiseless and noisy reconstructed cases (see Figs.~\ref{fig:noiseless_PDF},~\ref{fig:noiseless_pk},~\ref{fig:noisyPDF}, and \ref{fig:noisypk}). For AdvACT, we forecast the detection of non-Gaussianity to be $\approx$ 9$\sigma$ (PDF) and 6$\sigma$ (peak counts), after accounting for the non-Gaussianity of the reconstruction noise itself.  The non-Gaussianity of the noise has been neglected in previous estimates, but we show that it is non-negligible (Fig.~\ref{fig:recon}).

\item We confirm that current analytic forecasts for CMB lensing power spectrum constraints are accurate when confronted with constraints derived from our $N$-body pipeline that include the full non-Gaussian covariance (Fig.~\ref{fig:contour_fisher}).

\item An improvement of $\approx 30\%$ in the forecasted $\Omega_m$-$\sigma_8$ error contour is seen when the power spectrum is combined with PDF and peak counts (assuming AdvACT-level noise), compared to using the power spectrum alone. The covariance between the power spectrum and the other two non-Gaussian statistics is relatively small (with cross-covariance $< 20\%$ of the diagonal components), meaning the latter is complementary to the power spectrum.

\item For noiseless $\kappa$ maps (i.e., ignoring primordial CMB fluctuations and instrumental/atmospheric noise), a smaller smoothing kernel can help extract the most information from the PDF and peak counts (Fig.~\ref{fig:contour_noiseless}). For example, peak counts of 1.0 arcmin Gaussian smoothed maps alone can provide equally tight constraints as from the power spectrum.

\item We find non-zero skewness in the PDF and peak counts of reconstructed GRFs, which is absent from the input noiseless GRFs by definition. This skewness is the result of the quadratic estimator used for CMB lensing reconstruction from the temperature or polarization maps. Future forecasts for non-Gaussian CMB lensing statistics should include these effects, as we have here, or else the expected signal-to-noise could be overestimated.
\end{enumerate}

In this work, we have only considered temperature-based reconstruction estimators, but in the near future polarization-based estimators will have equally (and, eventually, higher) signal-to-noise.  Moreover, the polarization estimators allow the lensing field to be mapped out to smaller scales,
which suggests that they could be even more useful for non-Gaussian statistics.

In summary, there is rich information in CMB lensing maps that is not captured by two-point statistics, especially on small scales where nonlinear evolution is significant. In order to extract this information from future data from ongoing CMB Stage-III and near-future Stage-IV surveys, such as AdvACT, SPT-3G~\cite{Benson2014}, Simons Observatory\footnote{\url{http://www.simonsobservatory.org/} }, and CMB-S4~\cite{Abazajian2015}, non-Gaussian statistics must be studied and modeled carefully.  We have shown that non-Gaussian statistics will already contain useful information for Stage-III surveys, which suggests that their role in Stage-IV analyses will be even more important.  The payoff of these efforts could be significant, such as a quicker route to a neutrino mass detection.

\begin{acknowledgments}
We thank Nick Battaglia, Francois Bouchet, Simone Ferraro, Antony Lewis,  Mark Neyrinck, Emmanuel Schaan, and Marcel Schmittfull for useful discussions. We acknowledge helpful comments from an anonymous referee.
JL is supported by an NSF Astronomy and Astrophysics Postdoctoral Fellowship under award AST-1602663. This work is partially supported by a Junior Fellowship from the Simons Foundation to JCH and a Simons Fellowship to ZH. BDS is supported by a Fellowship from the Miller Institute for Basic Research in Science at the University of California, Berkeley. This work is partially supported by NSF grant AST-1210877 (to ZH) and by a ROADS award at Columbia University. This work used the Extreme Science and Engineering Discovery Environment (XSEDE), which is supported by NSF grant ACI-1053575. Computations were performed on the GPC supercomputer at the SciNet HPC consortium.  SciNet is funded by the Canada Foundation for Innovation under the auspices of Compute Canada, the Government of Ontario, the Ontario Research Fund --- Research Excellence, and the Univ.~of Toronto.
\end{acknowledgments}

\bibliographystyle{physrev}
\bibliography{paper}
\end{document}